\documentstyle[prd,aps,preprint,tighten,epsfig]{revtex}

\begin{document}

\draft

\title{A Systematic Study of Neutrino Mixing and CP Violation
from Lepton Mass Matrices with Six Texture Zeros}
\author{{\bf Shun Zhou} ~ and ~ {\bf Zhi-zhong Xing}}
\address{CCAST (World Laboratory), P.O. Box 8730, Beijing 100080, China \\
and Institute of High Energy Physics, Chinese Academy of Sciences, \\
P.O. Box 918 (4), Beijing 100049, China \\
({\it Electronic address: xingzz@mail.ihep.ac.cn})}
\maketitle

\begin{abstract}
We present a systematic study of 400 combinations of the
charged lepton and neutrino mass matrices with six vanishing
entries or texture zeros. Only 24 of them, which can be
classified into a few distinct categories, are found to be
compatible with current neutrino oscillation data at the
$3\sigma$ level. A peculiar feature of the lepton mass matrices
in each category is that they have the same phenomenological
consequences. Taking account of a simple seesaw
scenario for six parallel patterns of the charged lepton
and Dirac neutrino mass matrices with six zeros, we show
that it is possible to fit the experimental data at or below
the $2\sigma$ level. In particular, the maximal atmospheric
neutrino mixing can be reconciled with a strong neutrino mass
hierarchy in the seesaw case. Numerical predictions are also
obtained for the neutrino mass spectrum, flavor mixing angles,
CP-violating phases and effective masses of the tritium beta
decay and the neutrinoless double beta decay.
\end{abstract}


\newpage

\section{Introduction}
Recent solar \cite{SNO}, atmospheric \cite{SK}, reactor
(KamLAND \cite{KM} and CHOOZ \cite{CHOOZ}) and accelerator (K2K \cite{K2K})
neutrino oscillation experiments have provided us with
very convincing evidence that neutrinos are massive and lepton flavors
are mixed. In the framework of three lepton families, a full description
of the lepton mass spectra and flavor mixing at low energies needs twelve
physical parameters:
\begin{itemize}
\item     three charged lepton masses $m_e$, $m_\mu$ and $m_\tau$,
which have precisely been measured \cite{PDG};
\item     three neutrino masses $m_1$, $m_2$ and $m_3$, whose
relative sizes (i.e., two independent mass-squared differences
$\Delta m^2_{21} \equiv m^2_2 - m^2_1$ and
$\Delta m^2_{31} \equiv m^2_3 - m^2_1$) have roughly been known from
solar ($\Delta m^2_{21} \sim 10^{-5} ~ {\rm eV}^2$) and atmospheric
($|\Delta m^2_{31}| \sim 10^{-3} ~ {\rm eV}^2$) neutrino oscillations;
\item     three flavor mixing angles $\theta_{12}$, $\theta_{23}$ and
$\theta_{13}$, whose values have been determined or constrained to an
acceptable degree of accuracy from solar ($\theta_{12} \sim 33^\circ$),
atmospheric ($\theta_{23} \sim 45^\circ$) and reactor
($\theta_{13} < 13^\circ$) neutrino oscillations;
\item     three CP-violating phases $\delta$, $\rho$ and $\sigma$,
which are completely unrestricted by current neutrino data.
\end{itemize}
The future neutrino oscillation experiments are expected to
fix the sign of $\Delta m^2_{31}$, to pin down the magnitude of
$\theta_{13}$ and to probe the ``Dirac-type'' CP-violating phase
$\delta$. The proposed precision experiments
for the tritium beta decay and the neutrinoless double beta
decay will help determine or constrain the absolute scale of three
neutrino masses. Some information about the ``Majorana-type''
CP-violating phases $\rho$ and $\sigma$ may also be achieved from
a delicate measurement of the neutrinoless double beta decay.
However, it seems hopeless to separately determine $\rho$ and
$\sigma$ from any conceivable sets of feasible neutrino experiments
in the foreseeable future.

The phenomenology of lepton masses and flavor mixing at low
energies can be formulated in terms of the charged lepton mass matrix
$M_l$ and the (effective) neutrino mass matrix $M_\nu$. While the
former is in general arbitrary, the latter must be a symmetric matrix
required by the Majorana nature of three neutrino fields. Hence we
diagonalize $M_l$ by using two unitary matrices and $M_\nu$ by means of
a single unitary matrix:
\begin{eqnarray}
&& U^\dagger_l M_l \hat{U}_l \; = \left (\matrix{
m_e & 0 & 0 \cr
0 & m_\mu & 0 \cr
0 & 0 & m_\tau \cr} \right ) \; ,
\nonumber \\
&& U^\dagger_\nu M_\nu U^*_\nu = \left (\matrix{
m_1 & 0 & 0 \cr
0 & m_2 & 0 \cr
0 & 0 & m_3 \cr} \right ) \; .
%        (1)
\end{eqnarray}
The lepton flavor mixing matrix $V$ is defined as
$V \equiv U^\dagger_l U_\nu$, which describes the mismatch between
the diagonalizations of $M_l$ and $M_\nu$. In the flavor basis where
$M_l$ is diagonal and positive, $V$ directly links the neutrino
mass eigenstates $(\nu_1, \nu_2, \nu_3)$ to the neutrino flavor
eigenstates $(\nu_e, \nu_\mu, \nu_\tau)$:
\begin{equation}
\left ( \matrix{
\nu_e \cr
\nu_\mu \cr
\nu_\tau \cr} \right ) = \left ( \matrix{
V_{e1} & V_{e2} & V_{e3} \cr
V_{\mu 1} & V_{\mu 2} & V_{\mu 3} \cr
V_{\tau 1} & V_{\tau 2} & V_{\tau 3} \cr} \right )
\left ( \matrix{
\nu_1 \cr
\nu_2 \cr
\nu_3 \cr} \right ) \; .
%       (2)
\end{equation}
A convenient parametrization of $V$ is
\begin{equation}
V = \left ( \matrix{
c_{12} c_{13} & s_{12} c_{13} & s_{13} \cr
- c_{12} s_{23} s_{13} - s_{12} c_{23} e^{-i\delta} &
- s_{12} s_{23} s_{13} + c_{12} c_{23} e^{-i\delta} &
s_{23} c_{13} \cr
- c_{12} c_{23} s_{13} + s_{12} s_{23} e^{-i\delta} &
- s_{12} c_{23} s_{13} - c_{12} s_{23} e^{-i\delta} &
c_{23} c_{13} \cr } \right )
\left ( \matrix{
e^{i\rho}   & 0 & 0 \cr
0   & e^{i\sigma}   & 0 \cr
0   & 0 & 1 \cr} \right ) \; ,
%       (3)
\end{equation}
where $c_{ij} \equiv \cos\theta_{ij}$ and $s_{ij} \equiv \sin\theta_{ij}$
(for $ij=12,23,13$). We have known that
$\theta_{23} > \theta_{12} > \theta_{13}$ holds, but how small
$\theta_{13}$ is remains an open question. A global analysis of current
neutrino oscillation data shows that $\theta_{13}$ is most likely to
lie in the range $4^\circ \leq \theta_{13} \leq 6^\circ$ \cite{FIT}.
In this case, we are left with a {\it bi-large} mixing pattern of $V$,
which is quite different from the {\it tri-small} mixing pattern of the
quark flavor mixing matrix.

To interpret the observed hierarchy of $\Delta m^2_{21}$ and
$\Delta m^2_{31}$ as well as the bi-large lepton flavor mixing pattern,
many phenomenological ans$\rm\ddot{a}$tze of lepton mass matrices have
been proposed in the literature \cite{Review}.
A very interesting category of the ans$\rm\ddot{a}$tze focus on the
vanishing entries or {\it texture zeros} of $M_l$ and $M_\nu$ in a
specific flavor basis, from which some nontrivial and testable relations
can be established between the flavor mixing parameters and the lepton
mass ratios. We argue that texture zeros of lepton mass matrices might
result from a kind of new (Abelian or non-Abelian) flavor symmetry beyond
the standard electroweak model. Such zeros may dynamically mean that the
corresponding matrix elements are sufficiently suppressed in comparison
with their neighboring counterparts. From a phenomenological point of
view, the study of possible texture zeros of $M_l$ and $M_\nu$
at low energies {\it do} make sense, because it ought to help reveal the
underlying structures of leptonic Yukawa couplings at a superhigh energy
scale.

The main purpose of this article is to analyze the six-zero textures of
$M_l$ and $M_\nu$ in a systematic way. To be specific, we take $M_l$
to be symmetric, just as $M_\nu$ is. This point is true in a number of
SO(10) grand unification models, in which the group symmetry itself may
dictate all fermion mass matrices to be symmetric \cite{SO10}.
Then a pair of off-diagonal texture
zeros in $M_l$ or $M_\nu$ can be counted as one zero. We further require
that each mass matrix contain three texture zeros, such that the
moduli of its three non-vanishing elements can fully be determined in
terms of its three mass eigenvalues
%%%%%%%%%%%%%%%%%%%%%%%%%%%%%%%%%%
\footnote{One may certainly consider the possibility that one mass
matrix contains two zeros and the other consists of four zeros. In
this case, the former loses the calculability -- namely,
its four independent moduli cannot completely be calculated in terms
of its three mass eigenvalues; and the latter causes the correlation
between one of its three mass eigenvalues with the other two -- this
kind of mass correlation is in general incompatible with the relevant
experimental data. One must reject the possibility that one mass
matrix consists of one zero and the other contains five zeros,
because the latter only has a single non-vanishing mass eigenvalue
and is in strong conflict with our current knowledge about the charged
lepton or neutrino masses. Therefore, we restrict ourselves to the
most interesting and feasible case: six texture zeros are equally
shared between $M_l$ and $M_\nu$.}.
%%%%%%%%%%%%%%%%%%%%%%%%%%%%%%%%%%
Because there exist 20 different patterns of $M_l$ or
$M_\nu$ with three texture zeros, we totally obtain
$20\times 20 = 400$ combinations of $M_l$ and $M_\nu$ with six
texture zeros. A careful analysis shows that only 24 of them,
which can be classified into a few distinct categories, are
consistent with current neutrino oscillation data at the $3\sigma$
level. We find that the lepton mass matrices in each category
have a peculiar feature: they do not have the same structures,
but their phenomenological consequences are exactly the same.
This {\it isomeric} character makes the six-zero textures of lepton
mass matrices especially interesting for model building.
It is noticed that those 24 patterns of $M_l$ and $M_\nu$ are
difficult to agree with today's experimental data at the $2\sigma$
level, mainly due to a potential tension between the smallness of
$\Delta m^2_{21}/|\Delta m^2_{31}|$ and the largeness of
$\sin^2 \theta_{23}$. Taking account of a very simple seesaw
scenario for six parallel patterns of the charged lepton
and Dirac neutrino mass matrices with six zeros, we demonstrate
that it is possible to fit the present neutrino data at or below
the $2\sigma$ level. In particular, the maximal atmospheric
neutrino mixing (i.e., $\sin^2 2\theta_{23} \approx 1$)
can be reconciled with a strong neutrino mass hierarchy in the
seesaw case. Specific numerical predictions are also obtained
for the neutrino mass spectrum, flavor mixing angles, CP-violating
phases and effective masses of the tritium beta decay and the
neutrinoless double beta decay.

The remaining part of this article is organized as follows. A
classification of the six-zero textures of lepton mass matrices is
presented in section II, where a few criteria to select the
phenomenologically favorable patterns of $M_l$ and $M_\nu$ are
also outlined. Section III is devoted to the analytical
and numerical calculations of 24 patterns of lepton mass
matrices with or without the structural parallelism between $M_l$
and $M_\nu$. A simple
application of the seesaw mechanism to the charged lepton and Dirac
neutrino mass matrices with six texture zeros is illustrated in
section IV. Finally, we summarize our main results in section V.

\section{A classification of the six-zero textures}

A symmetric lepton mass matrix $M$ (i.e., $M_l$ or $M_\nu$) has six
independent entries. If three of them are taken to be vanishing, we
totally arrive at
\begin{equation}
^6{\bf C}_3 = \frac{6!}{3! \left ( 6 - 3 \right )!} = 20 \;
%       (4)
\end{equation}
patterns, which are structurally different from one another. These
twenty patterns of $M$ can be classified into four categories:
\begin{enumerate}
\item       Three diagonal matrix elements of $M$ are all vanishing
(type 0):
\begin{equation}
M_0 = \left ( \matrix{
{\bf 0} & \times    & \times \cr
\times  & {\bf 0}   & \times \cr
\times  & \times    & {\bf 0} \cr} \right ) \; , ~~~~~~~
%   (5)
\end{equation}
where those non-vanishing entries are simply symbolized by $\times$'s.
\item       Two diagonal matrix elements of $M$ are vanishing (type I):
\begin{eqnarray}
&& M_{\rm I_1} = \left ( \matrix{
{\bf 0} & \times  & {\bf 0} \cr
\times  & {\bf 0} & \times \cr
{\bf 0} & \times  & \times \cr} \right ) \; , ~~~
M_{\rm I_2} = \left ( \matrix{
{\bf 0} & {\bf 0} & \times \cr
{\bf 0} & \times  & \times \cr
\times  & \times  & {\bf 0} \cr} \right ) \; , ~~~
M_{\rm I_3} = \left ( \matrix{
{\bf 0} & \times  & \times \cr
\times  & {\bf 0} & {\bf 0} \cr
\times  & {\bf 0} & \times \cr} \right ) \; , ~~~~~~
\nonumber \\
&& M_{\rm I_4} = \left ( \matrix{
{\bf 0} & \times  & \times \cr
\times  & \times  & {\bf 0} \cr
\times  & {\bf 0} & {\bf 0} \cr} \right ) \; , ~~~
M_{\rm I_5} = \left ( \matrix{
\times  & {\bf 0} & \times \cr
{\bf 0} & {\bf 0} & \times \cr
\times  & \times  & {\bf 0} \cr} \right ) \; , ~~~
M_{\rm I_6} = \left ( \matrix{
\times  & \times  & {\bf 0} \cr
\times  & {\bf 0} & \times  \cr
{\bf 0} & \times  & {\bf 0} \cr} \right ) \; , ~~~~~~
%   (6)
\end{eqnarray}
which are of rank three; and
\begin{equation}
M_{\rm I_7} = \left ( \matrix{
{\bf 0} & {\bf 0} & \times \cr
{\bf 0} & {\bf 0} & \times \cr
\times  & \times  & \times \cr} \right ) \; , ~~~
M_{\rm I_8} = \left ( \matrix{
{\bf 0} & \times  & {\bf 0} \cr
\times  & \times  & \times \cr
{\bf 0} & \times  & {\bf 0} \cr} \right ) \; , ~~~
M_{\rm I_9} = \left ( \matrix{
\times  & \times  & \times \cr
\times  & {\bf 0} & {\bf 0} \cr
\times  & {\bf 0} & {\bf 0} \cr} \right ) \; , ~~~~~~~
%   (7)
\end{equation}
which are of rank two.
\item       One diagonal matrix element of $M$ is vanishing (type II):
\begin{eqnarray}
&& M_{\rm II_1} = \left ( \matrix{
\times  & \times  & {\bf 0} \cr
\times  & {\bf 0} & {\bf 0} \cr
{\bf 0} & {\bf 0} & \times \cr} \right ) \; , ~~~
M_{\rm II_2} = \left ( \matrix{
\times  & {\bf 0} & \times \cr
{\bf 0} & \times  & {\bf 0} \cr
\times  & {\bf 0} & {\bf 0} \cr} \right ) \; , ~~~
M_{\rm II_3} = \left ( \matrix{
{\bf 0} & \times  & {\bf 0} \cr
\times  & \times  & {\bf 0} \cr
{\bf 0} & {\bf 0} & \times \cr} \right ) \; , ~~~~~
\nonumber \\
&& M_{\rm II_4} = \left ( \matrix{
{\bf 0} & {\bf 0} & \times \cr
{\bf 0} & \times  & {\bf 0} \cr
\times  & {\bf 0} & \times  \cr} \right ) \; , ~~~
M_{\rm II_5} = \left ( \matrix{
\times  & {\bf 0} & {\bf 0} \cr
{\bf 0} & \times  & \times \cr
{\bf 0} & \times  & {\bf 0} \cr} \right ) \; , ~~~
M_{\rm II_6} = \left ( \matrix{
\times  & {\bf 0} & {\bf 0} \cr
{\bf 0} & {\bf 0} & \times  \cr
{\bf 0} & \times  & \times  \cr} \right ) \; , ~~~~~
%   (8)
\end{eqnarray}
which are of rank three; and
\begin{equation}
M_{\rm II_7} = \left ( \matrix{
\times  & \times  & {\bf 0} \cr
\times  & \times  & {\bf 0} \cr
{\bf 0} & {\bf 0} & {\bf 0} \cr} \right ) \; , ~~~
M_{\rm II_8} = \left ( \matrix{
\times  & {\bf 0} & \times \cr
{\bf 0} & {\bf 0} & {\bf 0} \cr
\times  & {\bf 0} & \times  \cr} \right ) \; , ~~~
M_{\rm II_9} = \left ( \matrix{
{\bf 0} & {\bf 0} & {\bf 0} \cr
{\bf 0} & \times  & \times  \cr
{\bf 0} & \times  & \times  \cr} \right ) \; , ~~~~~~
%   (9)
\end{equation}
which are of rank two.
\item       Three diagonal matrix elements of $M$ are all
non-vanishing (type III):
\begin{equation}
M_{\rm III} = \left ( \matrix{
\times  & {\bf 0} & {\bf 0} \cr
{\bf 0} & \times  & {\bf 0} \cr
{\bf 0} & {\bf 0} & \times  \cr} \right ) \; . ~~~~~~~
%       (10)
\end{equation}
\end{enumerate}
We see that $M_0$ and $M_{\rm I_1}$ are the well-known
Zee \cite{Zee} and Fritzsch \cite{F78} patterns of fermion
mass matrices, respectively. Both of them are disfavored in
the quark sector \cite{RRR}. While the original Zee ansatz
is also problematic in describing lepton masses and flavor
mixing \cite{He}, the Fritzsch ansatz is found to be essentially
compatible with current neutrino oscillation data \cite{X02}.

Allowing the charged lepton or neutrino mass matrix to take one of
the above three-zero textures, we totally have $20\times 20 = 400$
combinations of $M_l$ and $M_\nu$. We find that 141 of them can
easily be ruled out. First, the pattern in Eq. (5) is not suitable
for $M_l$, because three charged leptons have a strong mass hierarchy
and the sum of their masses (i.e., the trace of $M_\nu$) cannot be zero.
Second, the rank-two patterns in Eqs. (7) and (9) are not suitable
for $M_l$, because the former must have one vanishing mass eigenvalue.
Third, $M_l$ and $M_\nu$ cannot simultaneously take the pattern in
Eq. (10), otherwise there would be no lepton flavor mixing. We are
therefore left with $(20-7) \times 20 - 1 = 259$ combinations of $M_l$
and $M_\nu$.

To pick out the phenomenologically favorable six-zero patterns of
lepton mass matrices from 259 combinations of $M_l$ and $M_\nu$, one
has to confront their concrete predictions for the lepton mass spectra
and flavor mixing angles with current neutrino oscillation data.
The strategies to do so are outlined below:
\begin{enumerate}
\item       For each combination of $M_l$ and $M_\nu$, we do the
diagonalization like Eq. (1). Because $M_l$ has been specified to
be symmetric, $\hat{U}_l = U^*_l$ must hold. The matrix elements
of $U_l$ can be given in terms of two mass ratios
($x^{~}_l \equiv m_e/m_\mu \approx 0.00484$ and
$y^{~}_l \equiv m_\mu/m_\tau \approx 0.0594$ \cite{PDG}) and
two irremovable phase parameters
%%%%%%%%%%%%%%%%%%%%%%%%%%%%%%%%
\footnote{Without loss of generality, we can always arrange one of the
three non-vanishing entries of $M_l$ (or $M_\nu$) to be positive.
We are then left with two free phase parameters in $M_l$ (or $M_\nu$).}.
%%%%%%%%%%%%%%%%%%%%%%%%%%%%%%%%
A similar treatment is
applicable for the neutrino sector. The ratio of two independent
neutrino mass-squared differences reads
\begin{equation}
R_\nu \equiv \left | \frac{\Delta m^2_{21}}{\Delta m^2_{31}} \right |
= y^2_\nu ~ \frac{1 - x^2_\nu}{|1 - x^2_\nu y^2_\nu |} \; , ~~
%   (11)
\end{equation}
where $x_\nu \equiv m_1/m_2$ and $y_\nu \equiv m_2/m_3$. Note that
$x_\nu < 1$ (i.e., $m_1 < m_2$) must hold, but it remains unclear
whether $y_\nu < 1$ (normal mass hierarchy) or $y_\nu > 1$ (inverted
mass hierarchy). The numerical results for $\Delta m^2_{21}$ and
$|\Delta m^2_{31}|$, which are obtained from a global analysis of
current neutrino oscillation data \cite{FIT2},
have been listed in Table I.
We are therefore able to figure out the allowed range of $R_\nu$.
\item       The lepton flavor mixing matrix $V = U^\dagger_l U_\nu$
can then be obtained. Its nine elements depend on four mass ratios
($x^{~}_l$, $y^{~}_l$, $x_\nu$ and $y_\nu$) and two irremovable phase
combinations, which will subsequently be denoted as $\alpha$ and $\beta$.
In the standard parametrization of $V$, as shown in Eq. (2), one has
\begin{eqnarray}
\sin^2\theta_{12} & = & \frac{|V_{e2}|^2}{1 - |V_{e3}|^2} \; , ~~~~~~~
\nonumber \\
\sin^2\theta_{23} & = & \frac{|V_{\mu 3}|^2}{1 - |V_{e3}|^2} \; , ~~~~~~~
\nonumber \\
\sin^2\theta_{13} & = & |V_{e3}|^2 \; .
%       (12)
\end{eqnarray}
The experimental results for $\sin^2 \theta_{12}$, $\sin^2 \theta_{23}$
and $\sin^2 \theta_{13}$ are also listed in Table I.
\item       With the help of current experimental data, we make use
of Eqs. (11) and (12) to look for the parameter space of each pattern
of lepton mass matrices. The relevant free parameters include two
neutrino mass ratios ($x_\nu$ and $y_\nu$) and two CP-violating
phases ($\alpha$ and $\beta$). The latter may in general vary between
0 and $2\pi$. In our numerical analysis the points of $x_\nu$,
$y_\nu$, $\alpha$ and $\beta$ will be generated by scanning their
possible ranges according to a flat random number distribution.
Thus the density of output points in the $(x_\nu, y_\nu)$ and
$(\alpha, \beta)$ plots will be a clear reflection of strong constraints,
imposed by the neutrino oscillation data and the model (or ansatz)
itself, on these parameters.
A combination of $M_l$ and $M_\nu$ will be rejected, if
its parameter space is found to be empty.
\end{enumerate}
Of course, whether the parameter space of a specific pattern of
lepton mass matrices is empty or not depends on the confidence levels
of relevant experimental data. We shall focus on the $2\sigma$ and
$3\sigma$ intervals
of $\Delta m^2_{21}$, $\Delta m^2_{31}$, $\sin^2 \theta_{12}$,
$\sin^2 \theta_{23}$ and $\sin^2 \theta_{13}$ given in Ref. \cite{FIT2}.
It is worth mentioning that a plain scan of the unknown
parameters $(x_\nu, y_\nu)$ and $(\alpha, \beta)$ is empirically
simple and conservative, provided the reasonable ranges of
$\Delta m^2_{21}$ {\it etc} have been fixed. In this approximation
the error bars of those observables need not be statistically treated.

Examining all 259 combinations of the charged lepton and neutrino
mass matrices is a lengthy but straightforward work. We find that only
24 of them, whose $M_l$ and $M_\nu$ both belong to type I given in
Eqs. (6) and (7), are compatible with current neutrino oscillation
data at the $3\sigma$ level. The detailed analytical and numerical
calculations of those 24 patterns will be presented in sections III
and IV.

Once the parameter space of a given pattern of lepton mass matrices
is fixed, one may obtain some predictions for the neutrino mass
spectrum and leptonic CP violation. For example, the absolute values
of three neutrino masses can be determined as follows:
\begin{eqnarray}
m_3 & = & \frac{1}{\sqrt{|1 - y^2_\nu|}}
\sqrt{\Delta m^2_{\rm atm}} \;\; ,
\nonumber \\
m_2 & = & \frac{y_\nu}{\sqrt{|1 - y^2_\nu|}}
\sqrt{\Delta m^2_{\rm atm}} \; = \; \frac{1}{\sqrt{1 - x^2_\nu}}
\sqrt{\Delta m^2_{\rm sun}} \;\; ,
\nonumber \\
m_1 & = & \frac{x_\nu}{\sqrt{1 - x^2_\nu}}
\sqrt{\Delta m^2_{\rm sun}} \;\; .
%   (13)
\end{eqnarray}
Three CP-violating phases in the standard parametrization of $V$
are also calculable. As for CP violation in neutrino-neutrino
or antineutrino-antineutrino oscillations,
its strength is measured by the Jarlskog invariant
$\cal J$ \cite{Jarlskog}. The definition of $\cal J$ reads
\begin{equation}
{\rm Im} \left ( V_{a i} V_{b j} V^*_{a j} V^*_{b i} \right ) \; =\;
{\cal J} \sum_{c, k}
\left ( \epsilon_{a b c} \epsilon_{ijk} \right ) \; ,
%   (14)
\end{equation}
where the subscripts $(a, b, c)$ and $(i, j, k)$ run respectively over
$(e, \mu, \tau)$ and $(1,2,3)$. The magnitude of $\cal J$ depends
on both $(x_\nu, y_\nu)$ and $(\alpha, \beta)$. If
$|{\cal J}| \sim 1\%$ is achievable, then leptonic CP- and T-violating
effects could be measured in a variety of long-baseline neutrino
oscillation experiments \cite{LBL} in the future.

In addition, interesting predictions can be achieved for the
effective mass of the tritium beta decay $\langle m\rangle_e$
and that of the neutrinoless double beta decay $\langle m\rangle_{ee}$:
\begin{eqnarray}
&& \langle m\rangle^2_e \equiv \sum^3_{i=1}
\left ( m^2_i |V_{ei}|^2 \right )
\; = \; m^2_3 \left (x^2_\nu y^2_\nu |V_{e1}|^2 + y^2_\nu |V_{e2}|^2 +
|V_{e3}|^2 \right ) \; ,
\nonumber \\
&& \langle m\rangle_{ee} \equiv \left | \sum^3_{i=1}
\left ( m_i V^2_{ei} \right ) \right |
\; = \; m_3 \left | x_\nu y_\nu V^2_{e1} + y_\nu V^2_{e2} +
V^2_{e3} \right | \; .
%       (15)
\end{eqnarray}
The present experimental upper bound on $\langle m\rangle_e$ is
$\langle m\rangle_e < 2.2$ eV \cite{PDG}, while the sensitivity of the
proposed KATRIN experiment is expected to reach
$\langle m\rangle_e \sim 0.3$ eV \cite{K}. In comparison,
the upper limit $\langle m\rangle_{ee} < 0.35$ eV has been set by the
Heidelberg-Moscow Collaboration \cite{HM} at the $90\%$ confidence level
%%%%%%%%%%%%%%%%%%%%%%%
\footnote{If the reported evidence for the existence of the neutrinoless
double beta decay \cite{KK} is taken into account, one has
$0.05 ~ {\rm eV} \leq \langle m\rangle_{ee} \leq 0.84 ~ {\rm eV}$ at
the $95\%$ confidence level.}.
%%%%%%%%%%%%%%%%%%%%%%%
The sensitivity of the next-generation experiments
for the neutrinoless double beta decay is possible to reach
$\langle m\rangle_{ee} \sim 10$ meV to 50 meV \cite{B}.

\section{Favored patterns of lepton mass matrices}

The 24 patterns of lepton mass matrices, which are found to be
compatible with current neutrino oscillation data at the $3\sigma$
level, all belong to the type-I textures listed in Eqs. (6) and (7).
To make our subsequent discussions more convenient and concrete,
we rewrite those type-I textures of $M_l$ or $M_\nu$ and list them
in Table II. Two comments are in order.
\begin{itemize}
\item       Each type-I texture of $M$ (i.e., $M_l$ or $M_\nu$) can be
decomposed into $M = P \overline{M} P^T$, where $P$ denotes a diagonal
phase matrix and $\overline{M}$ is a real mass matrix with three
positive non-vanishing elements. The diagonalization of $\overline{M}$
requires an orthogonal transformation:
\begin{equation}
O^\dagger \overline{M} O^* \; = \; \left ( \matrix{
\lambda_1 & 0 & 0 \cr
0 & \lambda_2 & 0 \cr
0 & 0 & \lambda_3 \cr} \right ) \; , ~~~~~~~
%   (16)
\end{equation}
where $\lambda_i$ (for $i=1,2,3$) stand for the physical masses
of charged leptons (i.e., $\lambda_{1,2,3} = m_{e,\mu,\tau}$)
or neutrinos (i.e., $\lambda_i = m_i$). Then the unitary matrix $U$
(i.e., $U_l$ or $U_\nu$) used to diagonalize $M$ takes the form
$U = PO$.
\item        Note that the matrix elements of $\overline{M}$ and $O$
can be determined in terms of $\lambda_i$. This calculability
allows us to express the rank-3 (or rank-2) patterns of $M$ in a
universal way, as shown in Table II. It turns out that the
relation
\begin{equation}
M_{\rm I_n} = E_{\rm n} M_{\rm I_1} E^T_{\rm n} \; ,
~~~ ({\rm n = 1, \cdot\cdot\cdot, 6}) \;  ~~~~~~~
%       (17)
\end{equation}
holds for those rank-3 textures, where
\begin{eqnarray}
&& E_1 = \left ( \matrix{
1 & 0 & 0 \cr
0 & 1 & 0 \cr
0 & 0 & 1 \cr} \right ) \; , ~~~
E_2 = \left ( \matrix{
1 & 0 & 0 \cr
0 & 0 & 1 \cr
0 & 1 & 0 \cr} \right ) \; , ~~~
E_3 = \left ( \matrix{
0 & 1 & 0 \cr
1 & 0 & 0 \cr
0 & 0 & 1 \cr} \right ) \; , ~~~~
\nonumber \\
&& E_4 = \left ( \matrix{
0 & 1 & 0 \cr
0 & 0 & 1 \cr
1 & 0 & 0 \cr} \right ) \; , ~~~
E_5 = \left ( \matrix{
0 & 0 & 1 \cr
1 & 0 & 0 \cr
0 & 1 & 0 \cr} \right ) \; , ~~~
E_6 = \left ( \matrix{
0 & 0 & 1 \cr
0 & 1 & 0 \cr
1 & 0 & 0 \cr} \right ) \; . ~~~~
%       (18)
\end{eqnarray}
As for three rank-2 textures, we have
\begin{equation}
M_{\rm I_7} = E_1 M_{\rm I_7} E^T_1 \; , ~~~
M_{\rm I_8} = E_4 M_{\rm I_7} E^T_4 \; , ~~~
M_{\rm I_9} = E_5 M_{\rm I_7} E^T_5 \; . ~~~~~~
%       (19)
\end{equation}
It is easy to check that $E_{\rm n}$ is a real orthogonal matrix; i.e.,
$E_{\rm n} E^T_{\rm n} = E^T_{\rm n} E_{\rm n} = E_1$ holds.
In addition, $E_4 = E_2 E_3 = E_3 E_6 = E_6 E_2$ and $E_5 = E^T_4$ hold.
\end{itemize}
Eqs. (17) and (19) will be useful to demonstrate the isomeric features
of a few categories of lepton mass matrices with six texture zeros, as
one can see later on.

\subsection{Six Parallel Patterns (rank-3)}

We have six parallel patterns of $M_l$ and $M_\nu$,
\begin{equation} ~~~~~~~~
\begin{tabular}{|c|c|c|c|c|c|c|} \hline
%------------------------------------------------
~ $M_l$ ~ & ~ $\rm I_1$ ~ & ~ $\rm I_2$ ~ & ~ $\rm I_3$ ~ & ~ $\rm I_4$ ~
& ~ $\rm I_5$ ~ & ~ $\rm I_6$ ~ \\ \hline
%-------------------------
~ $M_\nu$ ~ & ~ $\rm I_1$ ~ & ~ $\rm I_2$ ~ & ~ $\rm I_3$ ~ &
~ $\rm I_4$ ~ & ~ $\rm I_5$ ~ & ~ $\rm I_6$ ~ \\ \hline
%-------------------------------------------------
\end{tabular} \;\;\; ,
%       (20)
\end{equation}
which are compatible with current neutrino oscillation data at the
$3\sigma$ level. Given $M^{l,\nu}_{\rm I_1}$ being diagonalized by
the unitary matrix $U_{l,\nu}$,
$M^{l,\nu}_{\rm I_n}$ (for $\rm n > 1$) can then be diagonalized by
$E_{\rm n} U_{l,\nu}$ as a result of Eq. (17). The lepton flavor
mixing matrix derived from $M^l_{\rm I_n}$ and $M^\nu_{\rm I_n}$ is
found to be identical to $V = U^\dagger_l U_\nu$, which is derived
from $M^l_{\rm I_1}$ and $M^\nu_{\rm I_1}$:
\begin{equation}
V_{\rm n} \equiv (E_{\rm n} U_l)^\dagger (E_{\rm n} U_\nu)
= U^\dagger_l (E^T_{\rm n} E_{\rm n}) U_\nu = V \; .
%       (21)
\end{equation}
This simple relation implies that six parallel patterns of $M_l$ and
$M_\nu$ are {\it isomeric} -- namely, they are structurally
different from one another, but their predictions for
lepton masses and flavor mixing are exactly the same \cite{XZ}. It
is therefore enough for us to consider only one of the six patterns
in the subsequent discussions. With the help of Eq. (16), the moduli
of three non-vanishing elements of $M_l$ or $M_\nu$ are given by
\begin{eqnarray}
A & = & \lambda_3 \left ( 1 - y + xy \right ) \; ,
\nonumber \\
B & = & \lambda_3 \left [ \frac{y (1 - x) (1 - y) (1 + xy)}
{1 - y + xy} \right ]^{1/2} \; ,
\nonumber \\
C & = & \lambda_3 \left ( \frac{x y^2}{1 - y + xy} \right )^{1/2} \; ,
%   (22)
\end{eqnarray}
where the subscript ``$l$'' or ``$\nu$'' has been omitted for simplicity.
Furthermore, we obtain the matrix elements of $O$ in terms
of the mass ratios $x$ and $y$ (see Table II for the definition of
$a^{~}_i$, $b^{~}_i$ and $c^{~}_i$):
\begin{eqnarray}
a^{~}_1 & = & + \left [ \frac{1-y}{(1+x)(1-xy)(1-y+xy)} \right ]^{1/2} \; ,
\nonumber \\
a^{~}_2 & = & -i \left [ \frac{x(1+xy)}{(1+x)(1+y)(1-y+xy)} \right ]^{1/2} \; ,
\nonumber \\
a^{~}_3 & = & + \left [ \frac{xy^3 (1-x)}{(1-xy)(1+y)(1-y+xy)}
\right ]^{1/2} \; ;
\nonumber \\
b^{~}_1 & = & + \left [ \frac{x(1-y)}{(1+x)(1-xy)} \right ]^{1/2} \; ,
\nonumber \\
b^{~}_2 & = & +i \left [ \frac{1+xy}{(1+x)(1+y)} \right ]^{1/2} \; ,
\nonumber \\
b^{~}_3 & = & + \left [ \frac{y(1-x)}{(1-xy)(1+y)} \right ]^{1/2} \; ;
\nonumber \\
c^{~}_1 & = & - \left [ \frac{xy(1-x)(1+xy)}{(1+x)(1-xy)(1-y+xy)}
\right ]^{1/2} \; ,
\nonumber \\
c^{~}_2 & = & -i \left [ \frac{y(1-x)(1-y)}{(1+x)(1+y)(1-y+xy)}
\right ]^{1/2} \; ,
\nonumber \\
c^{~}_3 & = & + \left [ \frac{(1-y)(1+xy)}{(1-xy)(1+y)(1-y+xy)}
\right ]^{1/2} \; .
%   (23)
\end{eqnarray}
Note that $a^{~}_2$, $b^{~}_2$ and $c^{~}_2$ are
imaginary, and their nontrivial phases arise from a minus sign of the
determinant of $M$(i.e., ${\rm Det}(M) = - AC^2 e^{2i\varphi}$).
Because of $0 < x_\nu < 1$ extracted from the solar neutrino oscillation
data \cite{SNO}, we can obtain $0 < y_\nu < 1$ from Eq. (22) as required
by the positiveness of $A_\nu$, $B_\nu$ and $C_\nu$
%%%%%%%%%%%%%%%%%%%%%%%%%%%%%%%%%%
\footnote{Although $y_\nu >1$ is in principle allowed by rephasing the
non-vanishing elements of $M_\nu$, our numerical analysis indicates that
this possibility is actually incompatible with current experimental data.}.
%%%%%%%%%%%%%%%%%%%%%%%%%%%%%%%%%%
Hence the six isomeric patterns of lepton mass matrices under discussion
guarantee a normal neutrino mass spectrum.

Nine elements of the lepton flavor mixing matrix
$V = U^\dagger_l U_\nu = O^\dagger_l (P^\dagger_l P_\nu) O_\nu$
can explicitly be written as
\begin{equation}
V_{pq} \; = \; (a^{l}_p)^* a^\nu_q e^{i\alpha} +
(b^{l}_p)^* b^\nu_q e^{i \beta} + (c^{l}_p)^* c^\nu_q \; ,
%   (24)
\end{equation}
where the subscripts $p$ and $q$ run respectively over $(e, \mu, \tau)$
and $(1,2,3)$, and the phase parameters $\alpha$ and $\beta$ are defined
by $\alpha \equiv (\varphi^{~}_\nu - \varphi^{~}_l) - \beta$ and
$\beta \equiv (\phi^{~}_\nu - \phi^{~}_l)$. Note that $V$ consists of
four free parameters $x_\nu$, $y_\nu$, $\alpha$ and $\beta$. The
latter can be constrained, with the help of Eqs. (11) and (12), by using
the experimental data listed in Table I ($\Delta m^2_{31} > 0$ as
a consequences of $0 < y_\nu < 1$). Once the parameter space of
$(x_\nu, y_\nu)$ and $(\alpha, \beta)$ is fixed, one may quantitatively
determine the Jarlskog invariant $\cal J$ and three CP-violating phases
$(\delta, \rho, \sigma)$. It is also possible to determine the neutrino
mass spectrum and two effective masses $\langle m\rangle_e$ and
$\langle m\rangle_{ee}$ defined in Eq. (15). The results of our numerical
calculations are summarized in Figs. 1--3. Some discussions are in order.
\begin{enumerate}
\item       We have noticed that the parameter space of $(x_\nu, y_\nu)$
or $(\alpha, \beta)$ will be empty, if the best-fit values or the
$2\sigma$ intervals of $\Delta m^2_{21}$, $\Delta m^2_{31}$,
$\sin^2 \theta_{12}$, $\sin^2 \theta_{23}$ and $\sin^2 \theta_{13}$
are taken into account. This situation is due to a potential conflict
between the largeness of $\sin^2\theta_{23}$ and the smallness of
$R_\nu$, which cannot simultaneously be fulfilled for six parallel
patterns of $M_l$ and $M_\nu$ at or below the $2\sigma$ level.

\item       If the $3\sigma$ intervals of $\Delta m^2_{21}$,
$\Delta m^2_{31}$, $\sin^2 \theta_{12}$, $\sin^2 \theta_{23}$ and
$\sin^2 \theta_{13}$ are used, however, the consequences of $M_l$ and
$M_\nu$ on two neutrino mass-squared differences and three flavor mixing
angles can be compatible with current experimental data.
Fig. 1 shows the allowed parameter space of $(x_\nu, y_\nu)$ and
$(\alpha, \beta)$ at the $3\sigma$ level. We see that $\beta \sim \pi$
holds. This result is certainly consistent with the
previous observation \cite{X02}. Because of $y_\nu \sim 0.25$,
$m_3 \approx \sqrt{\Delta m^2_{31}}$ is a good approximation. The
neutrino mass spectrum can actually be determined to an acceptable
degree of accuracy by using Eq. (13). For instance, we obtain
$m_3 \approx (3.8 - 6.1) \times 10^{-2}$ eV,
$m_2 \approx (0.95 - 1.5) \times 10^{-2}$ eV and
$m_1 \approx (2.6 - 3.4) \times 10^{-3}$ eV, where $x_\nu \approx 1/3$
and $y_\nu \approx 1/4$ have typically been taken.

\item       Fig. 2 shows the outputs of $\sin^2 \theta_{12}$,
$\sin^2 \theta_{23}$ and $\sin^2 \theta_{13}$ versus $R_\nu$ at the
$3\sigma$ level. One may observe that the maximal atmospheric neutrino
mixing (i.e., $\sin^2\theta_{23} \approx 0.5$ or
$\sin^2 2\theta_{23} \approx 1$) cannot be achieved from
the isomeric lepton mass matrices under consideration. To be specific,
$\sin^2\theta_{23} < 0.40$ (or $\sin^2 2\theta_{23} < 0.96$)
holds in our ansatz. It is impossible
to get a larger value of $\sin^2\theta_{23}$ even if $R_\nu$
approaches its upper limit. In contrast, the output of $\sin^2\theta_{12}$
is favorable and has less dependence on $R_\nu$. One may also see that
only small values of $\sin^2\theta_{13}$ ($\leq 0.016$) are favored.
More precise experimental data on $\sin^2\theta_{23}$, $\sin^2\theta_{13}$
and $R_\nu$ will allow us to examine whether those parallel patterns
of lepton mass matrices with six texture zeros can really survive the
experimental test or not.

\item       Fig. 3 illustrates the results of two effective masses
$\langle m\rangle_{e}$ and $\langle m\rangle_{ee}$, three CP-violating
phases $(\delta, \rho, \sigma)$, and the Jarlskog invariant $\cal J$.
It is obvious that $\langle m\rangle_e \sim 10^{-2} ~ {\rm eV}$ for the
tritium beta decay and $\langle m\rangle_{ee} \sim 10^{-3} ~ {\rm eV}$ for
the neutrinoless double beta decay. Both of them are too small to be
experimentally accessible in the foreseeable future. We find that
the maximal magnitude of $\cal J$ is close to 0.015 around
$\delta \sim 3\pi/4$ (or $5\pi/4$). As for the Majorana phases $\rho$
and $\sigma$, the relation $(\rho - \sigma) \approx \pi/2$ holds.
This result is attributed to the fact that the matrix elements
$(a^\nu_2, b^\nu_2, c^\nu_2)$ of $U_\nu$ are all imaginary and they
give rise to an irremovable phase shift between $V_{p1}$ and $V_{p2}$
(for $p=e, \mu, \tau$) elements through Eq. (23). Such a phase difference
affects $\langle m\rangle_{ee}$, but it has nothing to do with
$\langle m\rangle_e$ and $\cal J$.
\end{enumerate}
To relax the potential tension between the smallness of $R_\nu$ and
the largeness of $\sin^2\theta_{23}$, we shall incorporate a simple
seesaw scenario in the six-zero textures of charged lepton and Dirac
neutrino mass matrices in section IV.

\subsection{Six Non-parallel Patterns (rank-3)}

The following six non-parallel patterns of $M_l$ and $M_\nu$,
\begin{equation} ~~~~~~~
\begin{tabular}{|c|c|c|c|c|c|c|} \hline
%------------------------------------------------
~ $M_l$ ~ & ~ $\rm I_1$ ~ & ~ $\rm I_2$ ~ & ~ $\rm I_3$ ~ & ~ $\rm I_4$ ~
& ~ $\rm I_5$ ~ & ~ $\rm I_6$ ~ \\ \hline
%-------------------------
~ $M_\nu$ ~ & ~ $\rm I_2$ ~ & ~ $\rm I_1$ ~ & ~ $\rm I_5$ ~ &
~ $\rm I_6$ ~ & ~ $\rm I_3$ ~ & ~ $\rm I_4$ ~ \\ \hline
%-------------------------------------------------
\end{tabular} \;\;\; ,
%       (25)
\end{equation}
in which $M_\nu$ is of rank 3, are found to be compatible with current
neutrino oscillation data at the $3\sigma$ level.
Given $M^l_{\rm I_1}$ and $M^\nu_{\rm I_2}$ being diagonalized
respectively by the unitary matrices $U_l$ and $E_2U_\nu$,
where $U_{l,\nu} = P_{l, \nu} O_{l, \nu}$ with $O_{l, \nu}$ being
simple functions of $x^{~}_{l, \nu}$ and $y^{~}_{l, \nu}$ as
already shown in Eq. (23), the corresponding flavor mixing matrix reads
\begin{equation}
V_{pq} \; = \; (a^{l}_p)^* a^\nu_q e^{i\alpha} +
(b^{l}_p)^* c^\nu_q e^{i \beta} + (c^{l}_p)^* b^\nu_q \; ,
%   (26)
\end{equation}
where the subscripts $p$ and $q$ run respectively over $(e, \mu, \tau)$
and $(1,2,3)$, the phase parameters $\alpha$ and $\beta$ are defined
by $\alpha \equiv (\varphi^{~}_\nu - \varphi^{~}_l) -
(2\phi^{~}_\nu - \phi_l)$ and $\beta \equiv -(\phi^{~}_\nu + \phi^{~}_l)$,
and an overall phase factor $e^{i\phi^{~}_\nu}$ has been omitted.
Taking account of the other five combinations of $M_l$ and $M_\nu$ in
Eq. (25), we notice that $M^l_{\rm I_n}$ (for $\rm n \neq 1$) and
$M^\nu_{\rm I_n}$ (for $\rm n\neq 2$) can be diagonalized by
$E_{\rm n} U_l$ and $(E_{\rm n} E^T_2)U_\nu$, respectively.
Because the relation
\begin{equation}
E^T_2 (E_1 E^T_2) = E^T_3 (E_5 E^T_2) = E^T_4 (E_6 E^T_2)
= E^T_5 (E_3 E^T_2) = E^T_6 (E_4 E^T_2) = E_1
%       (27)
\end{equation}
holds, Eq. (26) is universally valid for
all six patterns. They are therefore isomeric.

We do a numerical analysis of six non-parallel patterns of
$M_l$ and $M_\nu$ in Eq. (25). The parameter space
of $(x_\nu, y_\nu)$ or $(\alpha, \beta)$ is found to be acceptable,
when the $3\sigma$ intervals of $\Delta m^2_{21}$, $\Delta m^2_{31}$,
$\sin^2 \theta_{12}$, $\sin^2 \theta_{23}$ and $\sin^2 \theta_{13}$
are used. Our explicit results are summarized in Figs. 4--6.
Some brief discussions are in order.
\begin{enumerate}
\item       Fig. 4 shows the allowed parameter space of $(x_\nu, y_\nu)$
and $(\alpha, \beta)$ at the $3\sigma$ level. We see that
$\beta \sim 0$ (or $\beta \sim 2\pi$) holds, while $\alpha$ is essentially
unrestricted. Again, $m_3 \approx \sqrt{\Delta m^2_{31}}$ is a good
approximation. The neutrino mass spectrum can roughly be determined
by using Eq. (13). Note that $x_\nu \sim 0.7$ is marginally allowed
-- in this case, $m_1$ and $m_2$ are approximately of the same order.
\item       The outputs of $\sin^2 \theta_{12}$,
$\sin^2 \theta_{23}$ and $\sin^2 \theta_{13}$ versus $R_\nu$ are
illustrated in Fig. 5. We are unable to obtain the maximal atmospheric
neutrino mixing (i.e., $\sin^2\theta_{23} \approx 0.5$ or equivalently
$\sin^2 2\theta_{23} \approx 1$) from
the non-parallel patterns of lepton mass matrices under consideration.
Indeed, $\sin^2\theta_{23} > 0.60$ (or $\sin^2 2\theta_{23} < 0.96$)
holds in our ansatz. It is impossible to get a larger value of
$\sin^2 2\theta_{23}$ even if $R_\nu$ approaches its upper bound.
In comparison, the output of $\sin^2\theta_{12}$ is favorable
and has less dependence on $R_\nu$. Only small values of
$\sin^2\theta_{13}$ ($\leq 0.02$) are allowed.
\item       The numerical results for $\langle m\rangle_{ee}/m_3$ versus
$\langle m\rangle_{e}/m_3$, $\cal J$ versus $\delta$, and
$\sigma$ versus $\rho$ are shown in Fig. 6.
Both $\langle m\rangle_e \sim 10^{-2} ~ {\rm eV}$ and
$\langle m\rangle_{ee} \sim 10^{-3} ~ {\rm eV}$ are too small to be
observable. The maximal magnitude of $\cal J$ is
close to 0.02 around $\delta \sim \pm \pi/4$, and the relation
$(\sigma - \rho) \approx \pi/2$ holds for two Majorana phases
of CP violation.
\end{enumerate}
Comparing the parallel patterns of $M_{l,\nu}$ in Eq. (20)
with those non-parallel patterns of $M_{l,\nu}$ in Eq. (25), we find
that most of their phenomenological consequences are quite similar.
Therefore, it is experimentally difficult to distinguish between them.

\subsection{Twelve Non-parallel Patterns ($m_1 =0$)}

Current neutrino oscillation data cannot exclude the possibility that
the neutrino mass $m_1$ or $m_3$ vanishes. Hence $M_\nu$ is in principle
allowed to take the rank-2 textures ($M_{\rm I_7}$, $M_{\rm I_8}$
and $M_{\rm I_9}$) listed in Table II. After a careful analysis, we
find that there exist four groups of non-parallel patterns of
$M_l$ and $M_\nu$ with $m_1 =0$, which are compatible with the present
experimental data at the $3\sigma$ level:
$$
\begin{tabular}{|c|c|c|c|} \hline
%------------------------------------------------
~ $M_l$ ~ & ~ $\rm I_1$ ~ & ~ $\rm I_4$ ~ & ~ $\rm I_5$ ~ \\ \hline
%-------------------------
~ $M_\nu$ ~ & ~ $\rm I_7$ ~ & ~ $\rm I_8$ ~ & ~ $\rm I_9$ ~ \\ \hline
%-------------------------------------------------
\end{tabular} \;\;\; ,
\eqno{\rm (28a)}
%       (28a)
$$
$$
\begin{tabular}{|c|c|c|c|} \hline
%------------------------------------------------
~ $M_l$ ~ & ~ $\rm I_3$ ~ & ~ $\rm I_2$ ~ & ~ $\rm I_6$ ~ \\ \hline
%-------------------------
~ $M_\nu$ ~ & ~ $\rm I_7$ ~ & ~ $\rm I_8$ ~ & ~ $\rm I_9$ ~ \\ \hline
%-------------------------------------------------
\end{tabular} \;\;\; ,
\eqno{\rm (28b)}
%       (28b)
$$
$$
\begin{tabular}{|c|c|c|c|} \hline
%------------------------------------------------
~ $M_l$ ~ & ~ $\rm I_2$ ~ & ~ $\rm I_6$ ~ & ~ $\rm I_3$ ~ \\ \hline
%-------------------------
~ $M_\nu$ ~ & ~ $\rm I_7$ ~ & ~ $\rm I_8$ ~ & ~ $\rm I_9$ ~ \\ \hline
%-------------------------------------------------
\end{tabular} \;\;\; ,
\eqno{\rm (28c)}
%       (28c)
$$
$$
\begin{tabular}{|c|c|c|c|} \hline
%------------------------------------------------
~ $M_l$ ~ & ~ $\rm I_5$ ~ & ~ $\rm I_1$ ~ & ~ $\rm I_4$ ~ \\ \hline
%-------------------------
~ $M_\nu$ ~ & ~ $\rm I_7$ ~ & ~ $\rm I_8$ ~ & ~ $\rm I_9$ ~ \\ \hline
%-------------------------------------------------
\end{tabular} \;\;\; .
\eqno{\rm (28d)}
%       (28d)
$$
The possibility of $m_3 = 0$ has been ruled out. With the help of
Eqs. (18) and (19), it is easy to prove that three combinations of
$M_l$ and $M_\nu$ in each of the above four groups are isomeric.
For the charged leptons, the expressions of $(A_l, B_l, C_l)$ and
$(a^{~}_i, b^{~}_i, c^{~}_i)$ can be found in Eqs. (22) and (23).
As for the neutrinos, we obtain
\setcounter{equation}{28}
\begin{eqnarray}
\tilde{A}_\nu & = & m_3 \left ( 1 - y_\nu \right ) \; ,
\nonumber \\
\tilde{B}_\nu & = & m_3 \sqrt{y_\nu - z^2_\nu} \;\; ,
%       (29)
\end{eqnarray}
where $z_\nu \equiv \tilde{C}_\nu/m_3$. We see that it is impossible
to fix $\tilde{C}_\nu$ (or $\tilde{B}_\nu$) in terms of $m_i$, due to
the fact that ${\rm Det}(M_\nu) = 0$ holds. This freedom
will be removed, however, once the flavor mixing parameters derived
from $M_l$ and $M_\nu$ are confronted with the experimental data.
To see this point more clearly, we write out the explicit results of
nine elements of the lepton flavor mixing matrix $V$ for every group
of $M_l$ and $M_\nu$:
$$
V_{pq} = (a^{l}_p)^* \tilde{a}^\nu_q e^{i\alpha} + (b^{l}_p)^*
\tilde{b}^\nu_q e^{i \beta} + (c^{l}_p)^* \tilde{c}^\nu_q \;
\eqno{\rm (30a)}
%       (30a)
$$
with $\alpha \equiv \tilde{\phi}^{~}_\nu - (\varphi^{~}_l - \phi^{~}_l)$
and $\beta \equiv \tilde{\varphi}^{~}_\nu - \phi^{~}_l$
corresponding to Eq. (28a);
$$
V_{pq} = (b^{l}_p)^* \tilde{a}^\nu_q e^{i\alpha} + (a^{l}_p)^*
\tilde{b}^\nu_q e^{i \beta} + (c^{l}_p)^* \tilde{c}^\nu_q \;
\eqno{\rm (30b)}
%       (30b)
$$
with $\alpha \equiv \tilde{\phi}^{~}_\nu - \phi^{~}_l$ and
$\beta \equiv \tilde{\varphi}^{~}_\nu - (\varphi^{~}_l - \phi^{~}_l)$
corresponding to Eq. (28b);
$$
V_{pq} = (a^{l}_p)^* \tilde{a}^\nu_q e^{i\alpha} + (c^{l}_p)^*
\tilde{b}^\nu_q e^{i \beta} + (b^{l}_p)^* \tilde{c}^\nu_q \;
\eqno{\rm (30c)}
%       (30c)
$$
with $\alpha \equiv \tilde{\phi}^{~}_\nu - (\varphi^{~}_l - 2\phi^{~}_l)$
and $\beta \equiv \tilde{\varphi}^{~}_\nu + \phi^{~}_l$
corresponding to Eq. (28c); and
$$
V_{pq} = (c^{l}_p)^* \tilde{a}^\nu_q e^{i\alpha} + (a^{l}_p)^*
\tilde{b}^\nu_q e^{i \beta} + (b^{l}_p)^* \tilde{c}^\nu_q \;
\eqno{\rm (30d)}
%       (30d)
$$
with $\alpha \equiv \tilde{\phi}^{~}_\nu + \phi^{~}_l$ and
$\beta \equiv \tilde{\varphi}^{~}_\nu - (\varphi^{~}_l - 2\phi^{~}_l)$
corresponding to Eq. (28d), where
\setcounter{equation}{30}
\begin{eqnarray}
&~~~~& \tilde{a}^\nu_1 = - \frac{z_\nu}{\sqrt{y_\nu}} \; , ~~~~~~~~~
\tilde{a}^\nu_2 = i \frac{\sqrt{y_\nu - z^2_\nu}}
{\sqrt{y_\nu + y^2_\nu}} \; , ~~~~~~
\tilde{a}^\nu_3 = \frac{\sqrt{y_\nu - z^2_\nu}}{\sqrt{1 + y_\nu}} \; ;
\nonumber \\
&& \tilde{b}^\nu_1 = \frac{\sqrt{y_\nu - z^2_\nu}}{\sqrt{y_\nu}} \; , ~~~~~
\tilde{b}^\nu_2 = i \frac{z_\nu}{\sqrt{y_\nu + y^2_\nu}} \; , ~~~~~~
\tilde{b}^\nu_3 = \frac{z_\nu}{\sqrt{1 + y_\nu}} \; ;
\nonumber \\
&& \tilde{c}^\nu_1 = 0 \; , ~~~~~~~~~~~~~~~~
\tilde{c}^\nu_2 = -i \frac{\sqrt{y_\nu}}{\sqrt{1 + y_\nu}} \; , ~~~~~
\tilde{c}^\nu_3 = \frac{1}{\sqrt{1 + y_\nu}} \; .
%       (31)
\end{eqnarray}
In obtaining Eqs. (30c) and (30d), we have omitted an overall phase
factor $e^{-i\phi^{~}_l}$.

Note that the sum $|\tilde{a}^\nu_i|^2 + |\tilde{b}^\nu_i|^2$
(for $i=1,2,3$) is independent of the free parameter $z_\nu$. This
result implies that $V_{pq}$ in Eq. (30b) can be arranged to amount
to $V_{pq}$ in Eq. (30a). Indeed, the replacements
$z_\nu \Longleftrightarrow \sqrt{y_\nu - z^2_\nu}~$ and
$\alpha \Longleftrightarrow \beta$ (or equivalently
$\tilde{\phi}^{~}_\nu \Longleftrightarrow \tilde{\varphi}^{~}_\nu$)
allow us to transform
$(V_{p1}, V_{p2}, V_{p3})$ of Eq. (30a) into
$(-V_{p1}, V_{p2}, V_{p3})$ of Eq. (30b). The extra minus sign of
$V_{p1}$ appearing in such a transformation does not make any physical
sense, because it can be removed by redefining the phases of three
charged lepton fields. Thus we expect that Eqs. (30a) and (30b)
lead to identical results for lepton flavor mixing and CP violation.
One may show that Eqs. (30c) and (30d) result in the same lepton
flavor mixing and CP violation in a similar way. For this reason,
it is only needed to numerically analyze the non-parallel patterns
of $M_l$ and $M_\nu$ in Eqs. (28a) and (28c).

A numerical analysis indicates that the parameter space of
$(y_\nu, z_\nu)$ or $(\alpha, \beta)$ can be found, if the
$3\sigma$ intervals of $\Delta m^2_{21}$, $\Delta m^2_{31}$,
$\sin^2 \theta_{12}$, $\sin^2 \theta_{23}$ and $\sin^2 \theta_{13}$
are taken into account. Our results are summarized in Figs. 7--12.
Some comments are in order.
\begin{enumerate}
\item       The parameter space and predictions of $M_l$ and $M_\nu$
listed in Eq. (28a) are shown in Figs. 7--9. We see that
$\beta \sim \pi$ is favored but $\alpha \sim \pi$ is disfavored.
The neutrino mass spectrum has a clear hierarchy:
$x_\nu = 0$ and $y_\nu \sim 0.25$. The outputs of
$\sin^2 \theta_{12}$ and $\sin^2 \theta_{23}$ are well constrained,
and they seem to favor the corresponding experimental
lower bounds. Again, it is impossible to obtain the maximal atmospheric
neutrino mixing. We observe that large values of $\sin^2 \theta_{13}$,
more or less close to its experimental upper limit, are strongly
favored. This interesting feature makes the present ansatz
experimentally distinguishable from those given in Eqs. (20) and (25).
As a straightforward consequence of the normal neutrino mass hierarchy,
the results of $\langle m\rangle_e$ and $\langle m\rangle_{ee}$ are
both too small to be observable in the near future.
The maximal magnitude of $\cal J$ is close to 0.02 around
$|\delta| \sim \pm \pi/7$. As for the Majorana phases, we get
the relation $(\sigma - \rho) \approx \pi/2$ (or $-3\pi/2$).
\item       The parameter space of $M_l$ and $M_\nu$ in
Eq. (28b) can be obtained from Fig. 7 with the replacements
$z_\nu \Longleftrightarrow \sqrt{y_\nu - z^2_\nu}~$ and
$\alpha \Longleftrightarrow \beta$. Such replacements are actually
equivalent to $\tilde{B}_\nu \Longleftrightarrow \tilde{C}_\nu$ and
$\tilde{\phi}^{~}_\nu \Longleftrightarrow \tilde{\varphi}^{~}_\nu$
between $M_\nu$ in Eq. (28a) and its counterpart in Eq. (28b).
The phenomenological consequences of $M_l$ and $M_\nu$ in both
cases are identical, as already shown above.
\item       Figs. 10--12 show the allowed parameter space and
predictions of $M_l$ and $M_\nu$ listed in Eq. (28c). We see that
$\alpha \sim \pi$ and $\beta \sim 0$ (or $2\pi$) are essentially
favored. The neutrino mass hierarchy is quite similar to
that illustrated in Fig. 7. The output of $\sin^2 \theta_{23}$ seems
to favor the corresponding experimental upper bound, and the maximal
atmospheric neutrino mixing cannot be achieved. In comparison, the
outputs of $\sin^2 \theta_{12}$ and $\sin^2 \theta_{13}$ are favorable
and have less dependence on $R_\nu$. Note that the predictions of this
ansatz for $\langle m\rangle_{ee}$ and $\cal J$ may reach
$0.4 m_3$ (at $\langle m\rangle_e \sim 0.15 m_3$)
and $0.03$ (at $\delta \sim \pm 3\pi/4$), respectively. Both results
are apparently larger than those obtained above.
Again, the relation $(\sigma - \rho) \approx \pi/2$ (or
$-3\pi/2$) holds for two Majorana phases.
\item       The parameter space of $M_l$ and $M_\nu$ in
Eq. (28d) can be obtained from Fig. 10 with the replacements
$z_\nu \Longleftrightarrow \sqrt{y_\nu - z^2_\nu}~$ and
$\alpha \Longleftrightarrow \beta$. Their phenomenological consequences
are identical to those derived from $M_l$ and $M_\nu$ in Eq. (28c).
\end{enumerate}
The main unsatisfactory output of twelve non-parallel patterns of
$M_l$ and $M_\nu$, just like the one of six parallel
patterns of $M_l$ and $M_\nu$ in Eq. (20),
is that $\sin^2 2\theta_{23}$ cannot reach the experimentally-favored
maximal value. Whether this is really a problem remains to be seen,
especially after more accurate neutrino oscillation data are
accumulated in the near future.

\section{A seesaw ansatz of lepton mass matrices}

To illustrate, let us discuss a simple way to avoid
the potential tension between the smallness of $R_\nu$ and the
largeness of $\sin^2\theta_{23}$ arising from those parallel patterns
of $M_l$ and $M_\nu$ in Eq. (20). In this connection,
we take account of the Fukugita-Tanimoto-Yanagida hypothesis \cite{FTY}
together with the seesaw mechanism \cite{SS} -- namely,
the charged lepton mass matrix $M_l$ and the Dirac neutrino mass matrix
$M_{\rm D}$ may take one of the six parallel patterns,
while the right-handed Majorana neutrino mass matrix $M_{\rm R}$ takes
the form $M_{\rm R} = M_0 E_1$ with $M_0$ denoting a very large mass
scale and $E_1$ being the unity matrix given in Eq. (18).
Then the effective (left-handed) neutrino mass matrix $M_\nu$ reads as
\begin{equation}
M_\nu \; =\; M_{\rm D} M^{-1}_R M^T_{\rm D} \; =\;
\frac{M^2_{\rm D}}{M_0} \; .
%       (32)
\end{equation}
For simplicity, we further assume $M_{\rm D}$ to be real (i.e.,
$\phi^{~}_{\rm D} = \varphi^{~}_{\rm D} =0$). It turns out that the real
orthogonal transformation $U_{\rm D}$, which is defined to diagonalize
$M_{\rm D}$, can simultaneously diagonalize $M_\nu$:
\begin{equation}
U^T_{\rm D} M_\nu U_{\rm D} \; =\;
\frac{(U^T_{\rm D} M_{\rm D} U_{\rm D})^2}{M_0} \; =\;
\left ( \matrix{
m_1 & 0 & 0 \cr
0 & m_2 & 0 \cr
0 & 0 & m_3 \cr} \right ) \; ,
%       (33)
\end{equation}
where $m_i \equiv d^2_i/M_0$ with $d_i$ standing for the eigenvalues
of $M_{\rm D}$. In terms of the neutrino mass ratios
$x_\nu \equiv m_1/m_2 = (d_1/d_2)^2$ and
$y_\nu \equiv m_2/m_3 = (d_2/d_3)^2$, we obtain the
explicit expressions of nine matrix elements of $U_\nu = U_{\rm D}$:
\begin{eqnarray}
a^\nu_1 & = & + \left [ \frac{1-\sqrt{y_\nu}}
{(1+\sqrt{x_\nu})(1-\sqrt{x_\nu y_\nu})
(1-\sqrt{y_\nu}+\sqrt{x_\nu y_\nu})} \right ]^{1/2} \; ,
\nonumber \\
a^\nu_2 & = & - \left [ \frac{\sqrt{x_\nu}(1+\sqrt{x_\nu y_\nu})}
{(1+\sqrt{x_\nu})(1+\sqrt{y_\nu})
(1-\sqrt{y_\nu}+\sqrt{x_\nu y_\nu})} \right ]^{1/2} \; ,
\nonumber \\
a^\nu_3 & = & + \left [ \frac{y_\nu \sqrt{x_\nu y_\nu}
(1-\sqrt{x_\nu})}{(1-\sqrt{x_\nu y_\nu})(1+\sqrt{y_\nu})
(1-\sqrt{y_\nu}+\sqrt{x_\nu y_\nu})} \right ]^{1/2} \; ,
\nonumber \\
b^\nu_1 & = & + \left [ \frac{\sqrt{x_\nu}(1-\sqrt{y_\nu})}
{(1+\sqrt{x_\nu})(1-\sqrt{x_\nu y_\nu})} \right ]^{1/2} \; ,
\nonumber \\
b^\nu_2 & = & + \left [ \frac{1+\sqrt{x_\nu y_\nu}}
{(1+\sqrt{x_\nu})(1+\sqrt{y_\nu})} \right ]^{1/2} \; ,
\nonumber \\
b^\nu_3 & = & + \left [ \frac{\sqrt{y_\nu}(1-\sqrt{x_\nu})}
{(1-\sqrt{x_\nu y_\nu})(1+\sqrt{y_\nu})} \right ]^{1/2} \; ,
\nonumber \\
c^\nu_1 & = & - \left [ \frac{\sqrt{x_\nu y_\nu}(1-\sqrt{x_\nu})
(1+\sqrt{x_\nu y_\nu})}{(1+\sqrt{x_\nu})(1-\sqrt{x_\nu y_\nu})
(1-\sqrt{y_\nu}+\sqrt{x_\nu y_\nu})} \right ]^{1/2} \; ,
\nonumber \\
c^\nu_2 & = & - \left [ \frac{\sqrt{y_\nu}(1-\sqrt{x_\nu})
(1-\sqrt{y_\nu})}{(1+\sqrt{x_\nu})(1+\sqrt{y_\nu})
(1-\sqrt{y_\nu}+\sqrt{x_\nu y_\nu})} \right ]^{1/2} \; ,
\nonumber \\
c^\nu_3 & = & + \left [ \frac{(1-\sqrt{y_\nu})(1+\sqrt{x_\nu y_\nu})}
{(1-\sqrt{x_\nu y_\nu})(1+\sqrt{y_\nu})(1-\sqrt{y_\nu}+\sqrt{x_\nu y_\nu})}
\right ]^{1/2} \; .
%   (34)
\end{eqnarray}
The lepton flavor mixing matrix $V = U^\dagger_l U_\nu$ remains to
take the same form as Eq. (24), but the relevant phase parameters
are now defined as $\alpha \equiv -\varphi^{~}_l -\beta$ and
$\beta \equiv - \phi^{~}_l$.
Comparing between Eqs. (23) and (34), one can immediately find that
the magnitudes of $(\theta_{12}, \theta_{23}, \theta_{13})$ in the
non-seesaw case can be reproduced in the seesaw case with much smaller
values of $x_\nu$ and $y_\nu$. The latter will allow $R_\nu$ to be
more strongly suppressed. It is therefore possible to relax the
tension between the smallness of $R_\nu$ and the largeness
of $\sin^2\theta_{23}$ appearing in the non-seesaw case. A careful
numerical analysis of six seesaw-modified patterns of lepton
mass matrices {\it does} support this observation. The results of our
calculations are summarized as follows.
\begin{enumerate}
\item       We find that the new ansatz are compatible very well with
current neutrino oscillation data, even if the $2\sigma$ intervals of
$\Delta m^2_{21}$, $\Delta m^2_{31}$, $\sin^2 \theta_{12}$,
$\sin^2 \theta_{23}$ and $\sin^2 \theta_{13}$ are taken into account.
Hence it is unnecessary to do a similar analysis at the $3\sigma$
level. The parameter space of $(x_\nu, y_\nu)$ and $(\alpha, \beta)$
is illustrated in Fig. 13, where $x_\nu \sim y_\nu \sim 0.2$ and
$\beta \sim \pi$ hold approximately.
Again $m_3 \approx \sqrt{\Delta m^2_{31}}$ is a good approximation.
The values of three neutrino masses read explicitly as
$m_3 \approx (4.2 - 5.8) \times 10^{-2}$ eV,
$m_2 \approx (0.84 - 1.2) \times 10^{-2}$ eV and
$m_1 \approx (1.6 - 1.9) \times 10^{-3}$ eV, which are obtained by
taking $x_\nu \approx y_\nu \approx 0.2$.
\item       The outputs of $\sin^2 \theta_{12}$,
$\sin^2 \theta_{23}$ and $\sin^2 \theta_{13}$ versus $R_\nu$ are shown
in Fig. 14 at the $2\sigma$ level. One can see that the magnitude of
$\sin^2 \theta_{12}$ is essentially unconstrained. Now the maximal
atmospheric neutrino mixing (i.e., $\sin^2\theta_{23} \approx 0.5$ or
$\sin^2 2\theta_{23} \approx 1$) is achievable in the region of
$R_\nu \sim 0.036-0.047$. It is also possible to obtain
$\sin^2\theta_{13} \leq 0.035$, just below the experimental upper
bound \cite{CHOOZ}. If $\sin^2 2\theta_{13} \geq 0.02$ really holds,
the measurement of $\theta_{13}$ should be realizable in a future
reactor neutrino oscillation experiment \cite{T13}.
\item      Fig. 15 illustrates the numerical results of
$\langle m\rangle_e$, $\langle m\rangle_{ee}$, $\delta$, $\rho$,
$\sigma$ and $\cal J$. We obtain
$\langle m\rangle_e \sim 10^{-2} ~ {\rm eV}$ for the tritium beta
decay and $\langle m\rangle_{ee} \sim 10^{-3} ~ {\rm eV}$ for the
neutrinoless double beta decay -- both of them are too small to be
experimentally accessible in the near future.
We see that $|{\cal J}| \sim 0.025$ can be
obtained. Such a size of CP violation is expected to be measured
in the future long-baseline neutrino oscillation experiments.
As for the Majorana phases $\rho$ and $\sigma$, the relation
$\sigma \approx \rho$ holds. This result is easily understandable,
because $U_\nu$ is real in the seesaw case. It is worth mentioning
that the effective neutrino mass matrix $M_\nu$ does not persist in
the simple texture as $M_l$ has, thus the allowed ranges of $\delta$,
$\rho$ and $\sigma$ become smaller in the seesaw case than in
the non-seesaw case.
\end{enumerate}

It should be noted that the eigenvalues of $M_{\rm D}$ and the heavy
Majorana mass scale $M_0$ are not specified in the above analysis.
But one may obtain $|d_1/d_2| = \sqrt{x_\nu} \sim 0.4$ and
$|d_2/d_3| = \sqrt{y_\nu} \sim 0.4$. Such a weak hierarchy of
$(|d_1|, |d_2|, |d_3|)$ means that $M_{\rm D}$ cannot directly
be connected to the charged lepton mass matrix $M_l$, nor can it
be related to the up-type quark mass matrix ($M_{\rm u}$) or
its down-type counterpart ($M_{\rm d}$) in a simple way. If the
hypothesis $M_{\rm R} = M_0 E_1$ is rejected but the result
$U^T_\nu M_\nu U_\nu = {\rm Diag}\{m_1, m_2, m_3\}$ with
$U_\nu$ given by Eq. (34) is maintained, it will be possible to
determine the pattern of $M_{\rm R}$ by means of the inverted
seesaw formula
$M_{\rm R} = M^T_{\rm D} M^{-1}_\nu M_{\rm D}$ \cite{XZhang} and by
assuming a specific relation between $M_{\rm D}$ and $M_{\rm u}$.
For example, one may simply assume $M_{\rm D} = M_{\rm u}$ with
$M_{\rm u}$ taking the approximate Fritzsch form,
\begin{equation}
M_{\rm u} \; \sim \; \left ( \matrix{
{\bf 0} & \sqrt{m_u m_c} & {\bf 0} \cr
\sqrt{m_u m_c} & {\bf 0} & \sqrt{m_c m_t} \cr
{\bf 0} & \sqrt{m_c m_t} & m_t \cr} \right ) \; .
%       (35)
\end{equation}
Just for the purpose of illustration, we typically input
$x_\nu \sim y_\nu \sim 0.18$ as well as
$m_u/m_c \sim m_c/m_t \sim 0.0031$ and $m_t \approx 175$ GeV at
the electroweak scale \cite{PDG}. Then we arrive at
\begin{equation}
M_{\rm R} \; \sim \; 3.0 \times 10^{15} \times \left ( \matrix{
6.1 \times 10^{-8} & 1.2 \times 10^{-5} & 2.0 \times 10^{-4} \cr
1.2 \times 10^{-5} & 3.5 \times 10^{-3} & 5.9 \times 10^{-2} \cr
2.0 \times 10^{-4} & 5.9 \times 10^{-2} & {\bf 1} \cr} \right ) \;
%       (36)
\end{equation}
in unit of GeV. This order-of-magnitude estimate shows that the
scale of $M_{\rm R}$ is close to that of grand unified theories
$\Lambda_{\rm GUT} \sim 10^{16}$ GeV, but the texture of $M_{\rm R}$
and that of $M_{\rm D}$ (or $M_l$) have little similarity. It is
certainly a very nontrivial task to combine the seesaw mechanism
and those phenomenologically-favored patterns of lepton mass
matrices. In this sense, the simple scenarios discussed in
Refs. \cite{XZ,FTY} and in the present paper may serve as a helpful
example to give readers a ball-park feeling of the problem itself
and possible solutions to it.

Of course, a similar application of the seesaw mechanism to the
non-parallel patterns of lepton mass matrices is straightforward.
In this case, an enhancement of $\sin^2 2\theta_{23}$ up to its
maximal value can also be achieved.

\section{Summary}

To summarize, we have analyzed 400 combinations of the charged lepton
and neutrino mass matrices with six texture zeros in a systematic
way. Only 24 of them, including 6 parallel patterns and 18 non-parallel
patterns, are found to be compatible with current neutrino
oscillation data at the $3\sigma$ level. Those viable patterns of
lepton mass matrices can be classified into a few distinct categories.
The textures in each category are demonstrated to have the same
phenomenological consequences, such as the normal neutrino mass
hierarchy and the bi-large flavor mixing pattern. We have also
discussed a very simple way to incorporate the seesaw mechanism in
the charged lepton and Dirac neutrino mass matrices with six
texture zeros. We illustrate that there is no problem to fit current
experimental data even at the $2\sigma$ level in the seesaw case.
In particular, the maximal atmospheric neutrino mixing can naturally
be reconciled with a relatively strong neutrino mass hierarchy.
Our results for effective masses of the tritium beta decay and the
neutrinoless double beta decay are too small to be experimentally
accessible in both the seesaw and non-seesaw cases, but the strength
of CP violation can reach the percent level and might be detectable in
the upcoming long-baseline neutrino oscillation experiments.

We conclude that the peculiar feature of isomeric lepton mass matrices
with six texture zeros is very suggestive for model building. We
therefore look forward to seeing whether such simple phenomenological
ans$\rm\ddot{a}$tze can survive the more stringent experimental test
or not in the near future.

\vspace{0.5cm}

This work was supported
in part by the National Natural Science Foundation of China.

\newpage

%%%%%%%%%%%%%%%%%%%%% Table 1 %%%%%%%%%%%%%%%%%%%%%%%
\begin{table}
\caption{The best-fit values, $2\sigma$ and $3\sigma$ intervals
of $\Delta m^2_{21}$, $|\Delta m^2_{31}|$, $\sin^2 \theta_{12}$,
$\sin^2 \theta_{23}$ and $\sin^2 \theta_{13}$ obtained from a
global analysis of the latest solar, atmospheric, reactor and
accelerator neutrino oscillation data [14].}
\begin{center}
\begin{tabular}{c|ccccc}
%----------------------------------------------------------------------
& $\Delta m^2_{21}$ ($10^{-5} ~ {\rm eV}^2$)
& $|\Delta m^2_{31}|$ ($10^{-3} ~ {\rm eV}^2$)
& $\sin^2 \theta_{12}$
& $\sin^2 \theta_{23}$
& $\sin^2 \theta_{13}$ \\ \hline
%-----------------------------------------------------
Best fit & 6.9 & 2.6 & 0.30 & 0.52 & 0.006 \cr
2$\sigma$ & 6.0--8.4 & 1.8--3.3 & 0.25--0.36 & 0.36--0.67 & $\leq$ 0.035 \cr
3$\sigma$ & 5.4--9.5 & 1.4--3.7 & 0.23--0.39 & 0.31--0.72 & $\leq$ 0.054
%-------------------------------------------------
\end{tabular}
\end{center}
\end{table}
%%%%%%%%%%%%%%%%%%%%%%%%%%%%%%%%%%%%%%%%%%%%%%%%%

%%%%%%%%%%%%%%%%%%%%% Table 2 %%%%%%%%%%%%%%%%%%%%%%%
\begin{table}
\caption{The type-I textures of a symmetric lepton mass
matrix $M$ (i.e., $M_l$ or $M_\nu$) and the corresponding forms of
the phase matrix $P$ (i.e., $P_l$ or $P_\nu$) and the unitary
matrix $O$ (i.e., $O_l$ or $O_\nu$) used to diagonalize $M$,
in which $(A,B,C)$ or $(\tilde{A},\tilde{B},\tilde{C})$ are
defined to be real and positive.}
\begin{center}
\begin{tabular}{c|ccc}
%----------------------------------------------------------------------
Rank 3 & The mass matrix $M$ & The phase matrix $P$ &
The unitary matrix $O$ \\ \hline
%----------------------------------------------------------------------
$\rm I_1$ &
$\left ( \matrix{
{\bf 0} & Ce^{i\varphi} & {\bf 0} \cr
Ce^{i\varphi} & {\bf 0} & Be^{i\phi} \cr
{\bf 0} & Be^{i\phi} & A \cr} \right )$ &
$\left ( \matrix{
e^{i(\varphi -\phi)} & 0 & 0 \cr
0 & e^{i\phi} & 0 \cr
0 & 0 & 1 \cr} \right )$ &
$\left ( \matrix{
a^{~}_1 & a^{~}_2 & a^{~}_3 \cr
b^{~}_1 & b^{~}_2 & b^{~}_3 \cr
c^{~}_1 & c^{~}_2 & c^{~}_3 \cr} \right )$ \\ \hline
%--------------------------------------------------
$\rm I_2$ &
$\left ( \matrix{
{\bf 0} & {\bf 0} & Ce^{i\varphi} \cr
{\bf 0} & A & Be^{i\phi} \cr
Ce^{i\varphi} & Be^{i\phi} & {\bf 0} \cr} \right )$ &
$\left ( \matrix{
e^{i(\varphi -\phi)} & 0 & 0 \cr
0 & 1 & 0 \cr
0 & 0 & e^{i\phi} \cr} \right )$ &
$\left ( \matrix{
a^{~}_1 & a^{~}_2 & a^{~}_3 \cr
c^{~}_1 & c^{~}_2 & c^{~}_3 \cr
b^{~}_1 & b^{~}_2 & b^{~}_3 \cr} \right )$ \\ \hline
%--------------------------------------------------
$\rm I_3$ &
$\left ( \matrix{
{\bf 0} & Ce^{i\varphi} & Be^{i\phi} \cr
Ce^{i\varphi} & {\bf 0} & {\bf 0} \cr
Be^{i\phi} & {\bf 0} & A \cr} \right )$ &
$\left ( \matrix{
e^{i\phi} & 0 & 0 \cr
0 & e^{i(\varphi-\phi)} & 0 \cr
0 & 0 & 1 \cr} \right )$ &
$\left ( \matrix{
b^{~}_1 & b^{~}_2 & b^{~}_3 \cr
a^{~}_1 & a^{~}_2 & a^{~}_3 \cr
c^{~}_1 & c^{~}_2 & c^{~}_3 \cr} \right )$ \\ \hline
%--------------------------------------------------
$\rm I_4$ &
$\left ( \matrix{
{\bf 0} & Be^{i\phi} & Ce^{i\varphi} \cr
Be^{i\phi} & A & {\bf 0} \cr
Ce^{i\varphi} & {\bf 0} & {\bf 0} \cr} \right )$ &
$\left ( \matrix{
e^{i\phi} & 0 & 0 \cr
0 & 1 & 0 \cr
0 & 0 & e^{i(\varphi-\phi)} \cr} \right )$ &
$\left ( \matrix{
b^{~}_1 & b^{~}_2 & b^{~}_3 \cr
c^{~}_1 & c^{~}_2 & c^{~}_3 \cr
a^{~}_1 & a^{~}_2 & a^{~}_3 \cr} \right )$ \\ \hline
%--------------------------------------------------
$\rm I_5$ &
$\left ( \matrix{
A & {\bf 0} & Be^{i\phi} \cr
{\bf 0} & {\bf 0} & Ce^{i\varphi} \cr
Be^{i\phi} & Ce^{i\varphi} & {\bf 0} \cr} \right )$ &
$\left ( \matrix{
1 & 0 & 0 \cr
0 & e^{i(\varphi-\phi)} & 0 \cr
0 & 0 & e^{i\phi} \cr} \right )$ &
$\left ( \matrix{
c^{~}_1 & c^{~}_2 & c^{~}_3 \cr
a^{~}_1 & a^{~}_2 & a^{~}_3 \cr
b^{~}_1 & b^{~}_2 & b^{~}_3 \cr} \right )$ \\ \hline
%--------------------------------------------------
$\rm I_6$ &
$\left ( \matrix{
A & Be^{i\phi} & {\bf 0} \cr
Be^{i\phi} & {\bf 0} & Ce^{i\varphi} \cr
{\bf 0} & Ce^{i\varphi} & {\bf 0} \cr} \right )$ &
$\left ( \matrix{
1 & 0 & 0 \cr
0 & e^{i\phi} & 0 \cr
0 & 0 & e^{i(\varphi-\phi)} \cr} \right )$ &
$\left ( \matrix{
c^{~}_1 & c^{~}_2 & c^{~}_3 \cr
b^{~}_1 & b^{~}_2 & b^{~}_3 \cr
a^{~}_1 & a^{~}_2 & a^{~}_3 \cr} \right )$ \\ \hline\hline
%----------------------------------------------------------------------
Rank 2 & The mass matrix $M$ & The phase matrix $P$ &
The unitary matrix $O$ \\ \hline
%----------------------------------------------------------------------
$\rm I_7$ &
$\left ( \matrix{
{\bf 0} & {\bf 0} & \tilde{B}e^{i\tilde{\phi}} \cr
{\bf 0} & {\bf 0} & \tilde{C}e^{i\tilde{\varphi}} \cr
\tilde{B}e^{i\tilde{\phi}} & \tilde{C}e^{i\tilde{\varphi}}
& \tilde{A} \cr} \right )$ &
$\left ( \matrix{
e^{i\tilde{\phi}} & 0 & 0 \cr
0 & e^{i\tilde{\varphi}} & 0 \cr
0 & 0 & 1 \cr} \right )$ &
$\left ( \matrix{
\tilde{a}^{~}_1 & \tilde{a}^{~}_2 & \tilde{a}^{~}_3 \cr
\tilde{b}^{~}_1 & \tilde{b}^{~}_2 & \tilde{b}^{~}_3 \cr
\tilde{c}^{~}_1 & \tilde{c}^{~}_2 & \tilde{c}^{~}_3 \cr} \right )$ \\ \hline
%--------------------------------------------------
$\rm I_8$ &
$\left ( \matrix{
{\bf 0} & \tilde{C}e^{i\tilde{\varphi}} & {\bf 0} \cr
\tilde{C}e^{i\tilde{\varphi}} & \tilde{A}
& \tilde{B}e^{i\tilde{\phi}} \cr
{\bf 0} & \tilde{B}e^{i\tilde{\phi}} & {\bf 0} \cr} \right )$ &
$\left ( \matrix{
e^{i\tilde{\varphi}} & 0 & 0 \cr
0 & 1 & 0 \cr
0 & 0 & e^{i\tilde{\phi}} \cr} \right )$ &
$\left ( \matrix{
\tilde{b}^{~}_1 & \tilde{b}^{~}_2 & \tilde{b}^{~}_3 \cr
\tilde{c}^{~}_1 & \tilde{c}^{~}_2 & \tilde{c}^{~}_3 \cr
\tilde{a}^{~}_1 & \tilde{a}^{~}_2 & \tilde{a}^{~}_3 \cr} \right )$ \\ \hline
%--------------------------------------------------
$\rm I_9$ &
$\left ( \matrix{
\tilde{A} & \tilde{B}e^{i\tilde{\phi}}
& \tilde{C}e^{i\tilde{\varphi}} \cr
\tilde{B}e^{i\tilde{\phi}} & {\bf 0} & {\bf 0} \cr
\tilde{C}e^{i\tilde{\varphi}} & {\bf 0} & {\bf 0} \cr} \right )$ &
$\left ( \matrix{
1 & 0 & 0 \cr
0 & e^{i\tilde{\phi}} & 0 \cr
0 & 0 & e^{i\tilde{\varphi}} \cr} \right )$ &
$\left ( \matrix{
\tilde{c}^{~}_1 & \tilde{c}^{~}_2 & \tilde{c}^{~}_3 \cr
\tilde{a}^{~}_1 & \tilde{a}^{~}_2 & \tilde{a}^{~}_3 \cr
\tilde{b}^{~}_1 & \tilde{b}^{~}_2 & \tilde{b}^{~}_3 \cr} \right )$ \\
%-------------------------------------------------
\end{tabular}
\end{center}
\end{table}
%%%%%%%%%%%%%%%%%%%%%%%%%%%%%%%%%%%%%%%%%%%%%%%%%

\newpage

%%%%%%%%%%%%%%%%%%%% Fig. 1 %%%%%%%%%%%%%%%%
\begin{figure}[t]
\vspace{-4cm}
\epsfig{file=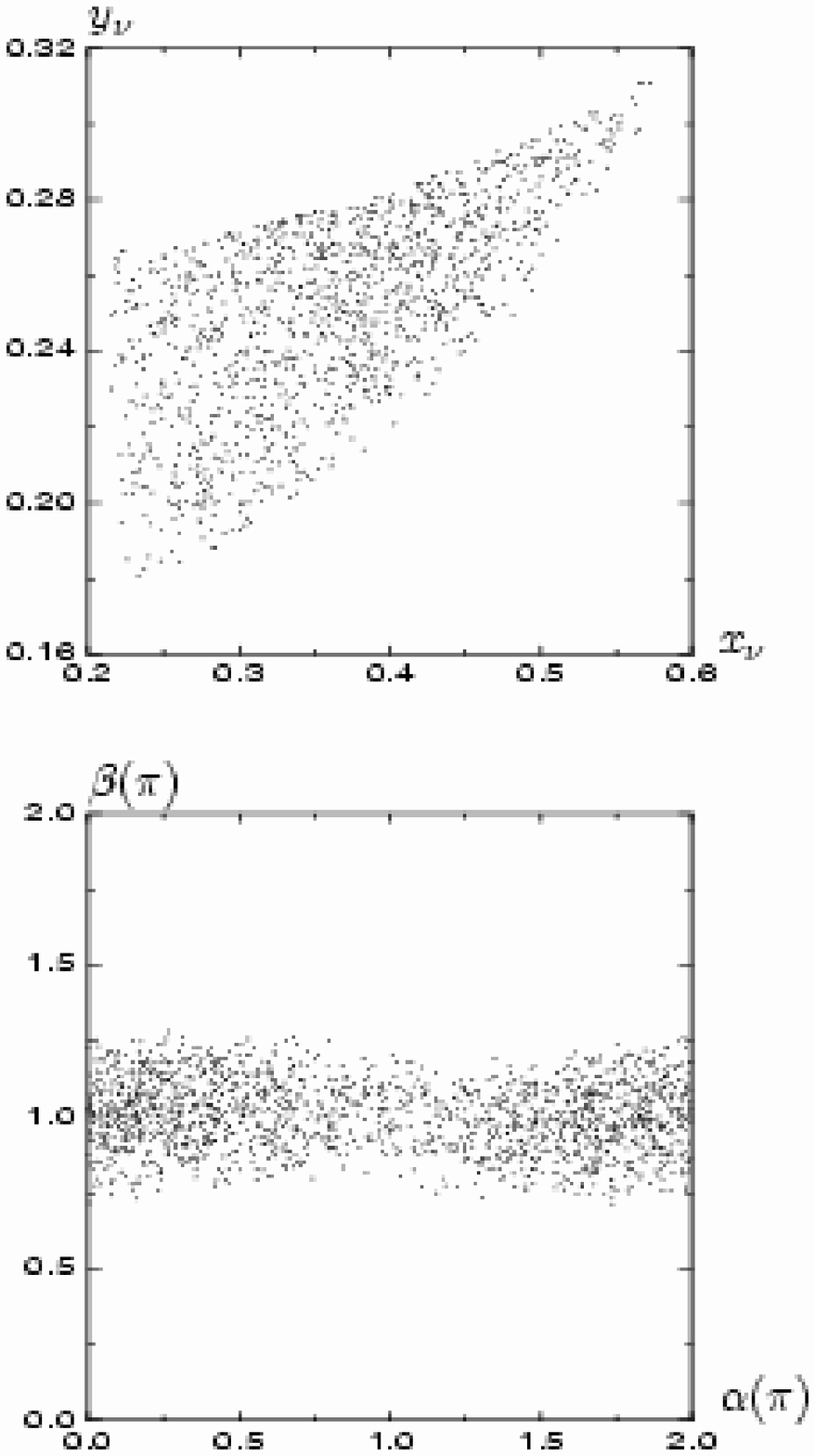,bbllx=-5.5cm,bblly=11cm,bburx=8cm,bbury=25cm,%
width=10cm,height=10cm,angle=0,clip=0} \vspace{12cm}
\caption{\underline{Parallel patterns of $M_l$ and $M_\nu$ in Eq.
(20):} the parameter space of $(x_\nu, y^{~}_\nu)$ and $(\alpha,
\beta)$ at the $3\sigma$ level.}
\end{figure}
%%%%%%%%%%%%%%%%%%%%%%%%%%%%%%%%%%%%%%%%%%%

%%%%%%%%%%%%%%%%%%%% Fig. 2 %%%%%%%%%%%%%%%%
\begin{figure}[t]
\vspace{0cm}
\epsfig{file=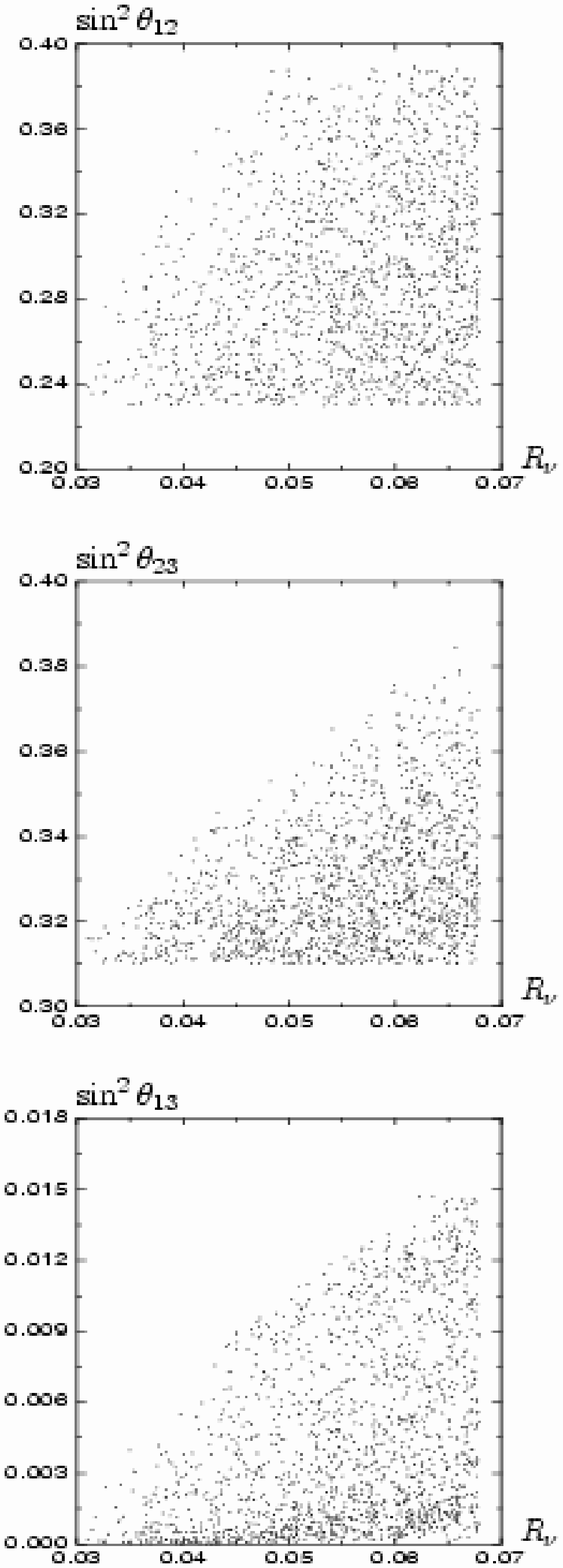,bbllx=-5cm,bblly=13cm,bburx=7cm,bbury=27cm,%
width=9cm,height=10cm,angle=0,clip=0} \vspace{11cm}
\caption{\underline{Parallel patterns of $M_l$ and $M_\nu$ in Eq.
(20):} the outputs of $\sin^2 \theta_{12}$, $\sin^2 \theta_{23}$
and $\sin^2 \theta_{13}$ versus $R_\nu$ at the $3\sigma$ level.}
\end{figure}
%%%%%%%%%%%%%%%%%%%%%%%%%%%%%%%%%%%%%%%%%%%

%%%%%%%%%%%%%%%%%%%% Fig. 3 %%%%%%%%%%%%%%%%
\begin{figure}[t]
\vspace{0cm}
\epsfig{file=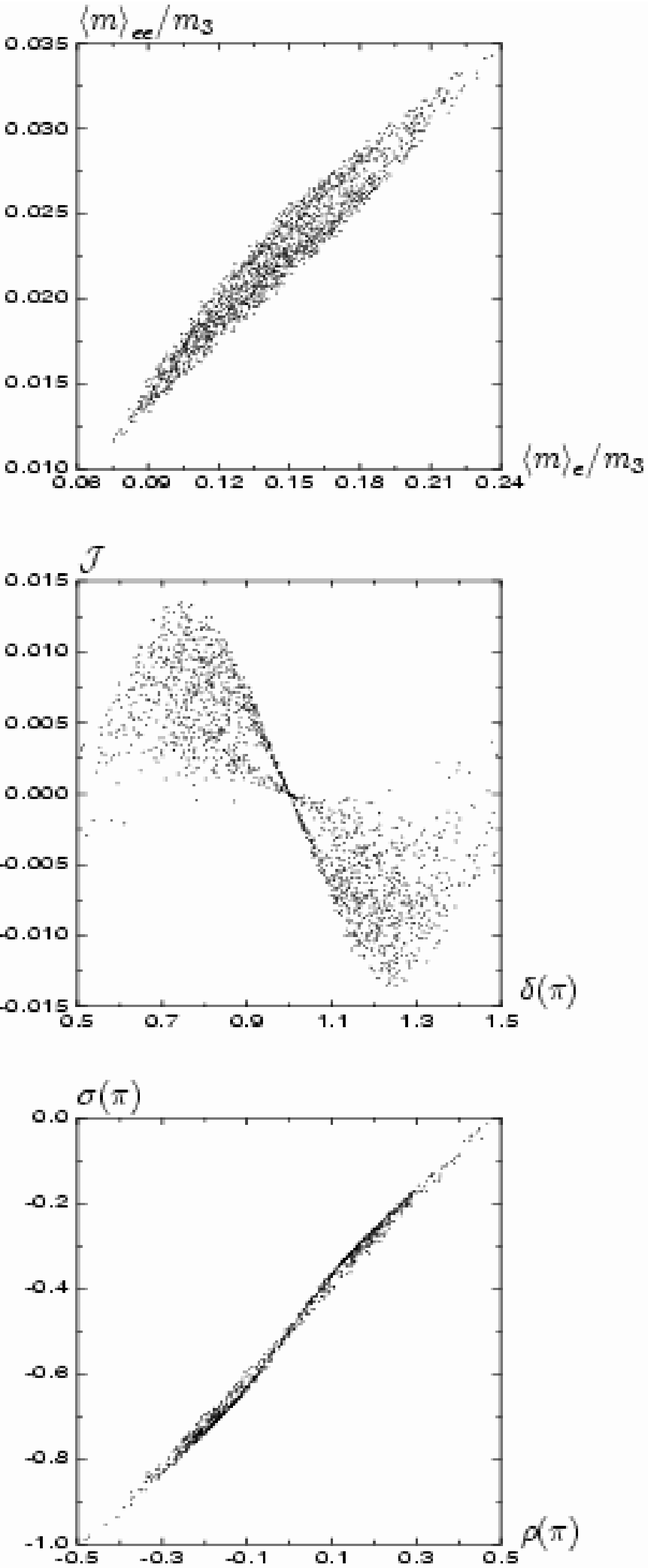,bbllx=-5cm,bblly=13cm,bburx=7cm,bbury=27cm,%
width=9cm,height=10cm,angle=0,clip=0} \vspace{11cm}
\caption{\underline{Parallel patterns of $M_l$ and $M_\nu$ in Eq.
(20):} the outputs of $(\langle m\rangle_e, \langle
m\rangle_{ee})$, $(\delta, {\cal J})$ and $(\rho, \sigma)$ at the
$3\sigma$ level.}
\end{figure}
%%%%%%%%%%%%%%%%%%%%%%%%%%%%%%%%%%%%%%%%%%%

%%%%%%%%%%%%%%%%%%%% Fig. 4 %%%%%%%%%%%%%%%%
\begin{figure}[t]
\vspace{-2cm}
\epsfig{file=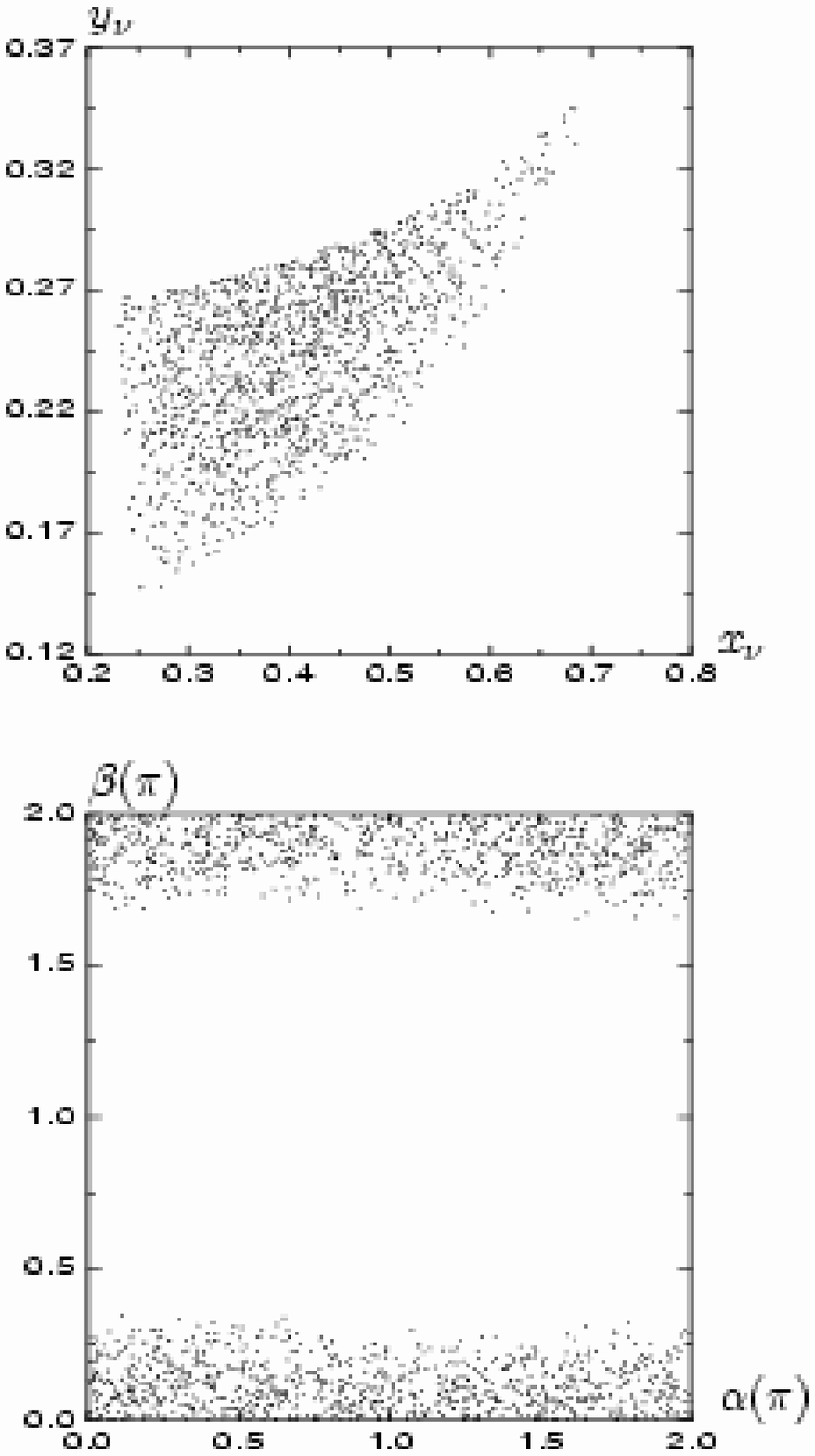,bbllx=-5.5cm,bblly=8cm,bburx=8cm,bbury=22cm,%
width=10cm,height=10cm,angle=0,clip=0} \vspace{8cm}
\caption{\underline{Non-parallel patterns of $M_l$ and $M_\nu$ in
Eq. (25):} the parameter space of $(x_\nu, y^{~}_\nu)$ and
$(\alpha, \beta)$ at the $3\sigma$ level.}
\end{figure}
%%%%%%%%%%%%%%%%%%%%%%%%%%%%%%%%%%%%%%%%%%%

%%%%%%%%%%%%%%%%%%%% Fig. 5 %%%%%%%%%%%%%%%%
\begin{figure}[t]
\vspace{0cm}
\epsfig{file=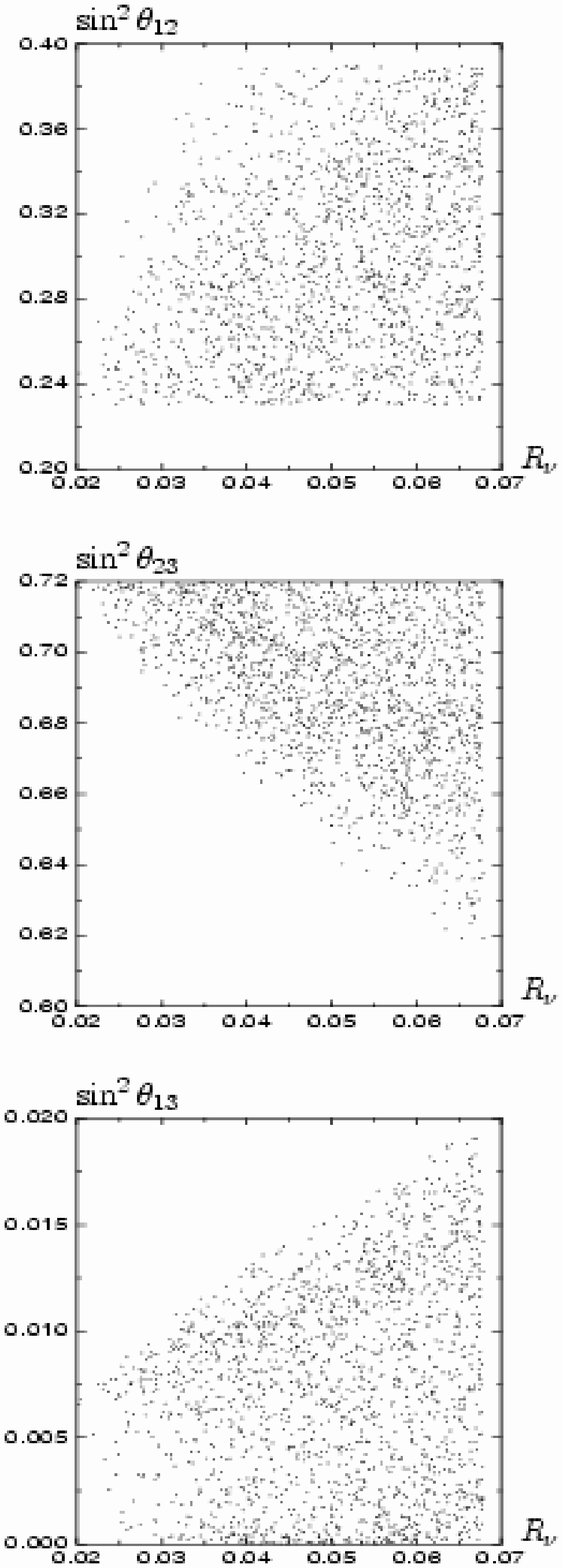,bbllx=-5cm,bblly=13cm,bburx=7cm,bbury=27cm,%
width=9cm,height=10cm,angle=0,clip=0} \vspace{11cm}
\caption{\underline{Non-parallel patterns of $M_l$ and $M_\nu$ in
Eq. (25):} the outputs of $\sin^2 \theta_{12}$, $\sin^2
\theta_{23}$ and $\sin^2 \theta_{13}$ versus $R_\nu$ at the
$3\sigma$ level.}
\end{figure}
%%%%%%%%%%%%%%%%%%%%%%%%%%%%%%%%%%%%%%%%%%%

%%%%%%%%%%%%%%%%%%%% Fig. 6 %%%%%%%%%%%%%%%%
\begin{figure}[t]
\vspace{0cm}
\epsfig{file=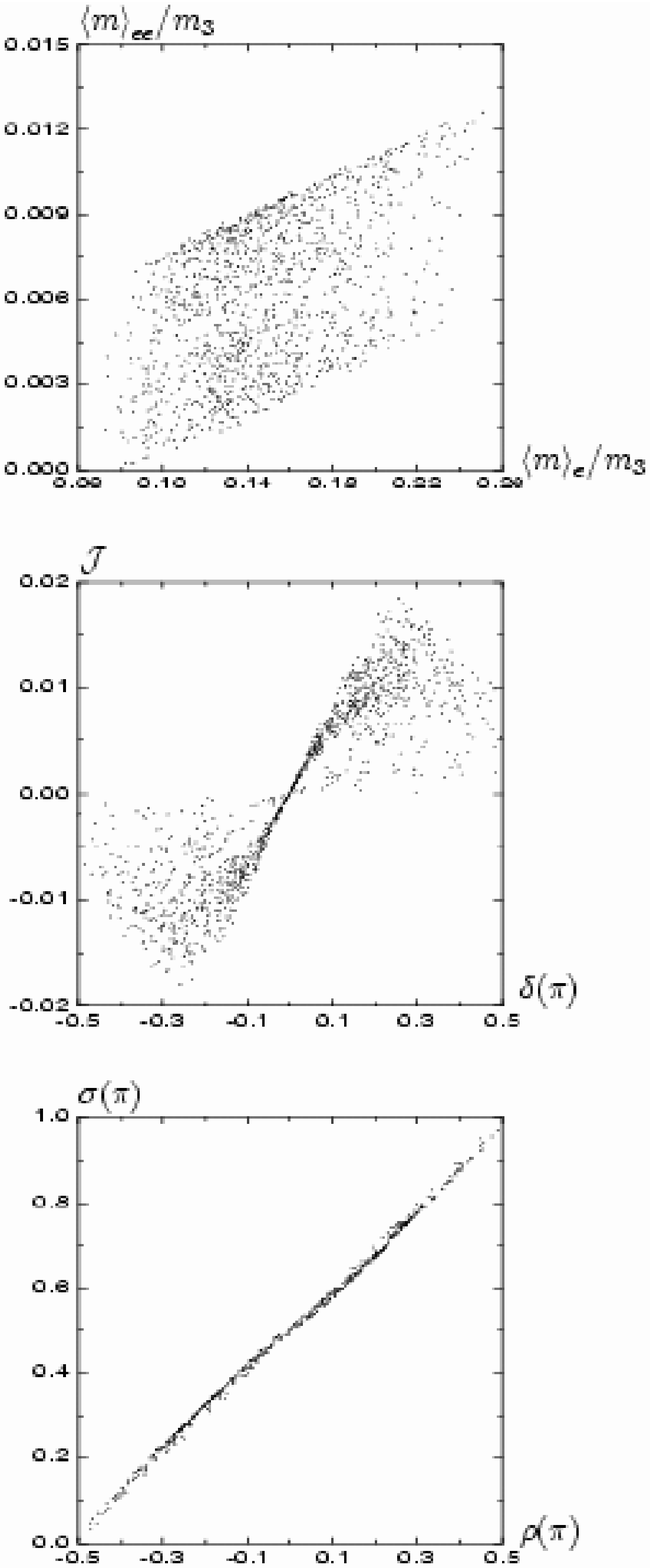,bbllx=-5cm,bblly=13cm,bburx=7cm,bbury=27cm,%
width=9cm,height=10cm,angle=0,clip=0} \vspace{11cm}
\caption{\underline{Non-parallel patterns of $M_l$ and $M_\nu$ in
Eq. (25):} the outputs of $(\langle m\rangle_e, \langle
m\rangle_{ee})$, $(\delta, {\cal J})$ and $(\rho, \sigma)$ at the
$3\sigma$ level.}
\end{figure}
%%%%%%%%%%%%%%%%%%%%%%%%%%%%%%%%%%%%%%%%%%%

%%%%%%%%%%%%%%%%%%%% Fig. 7 %%%%%%%%%%%%%%%%
\begin{figure}[t]
\vspace{-2cm}
\epsfig{file=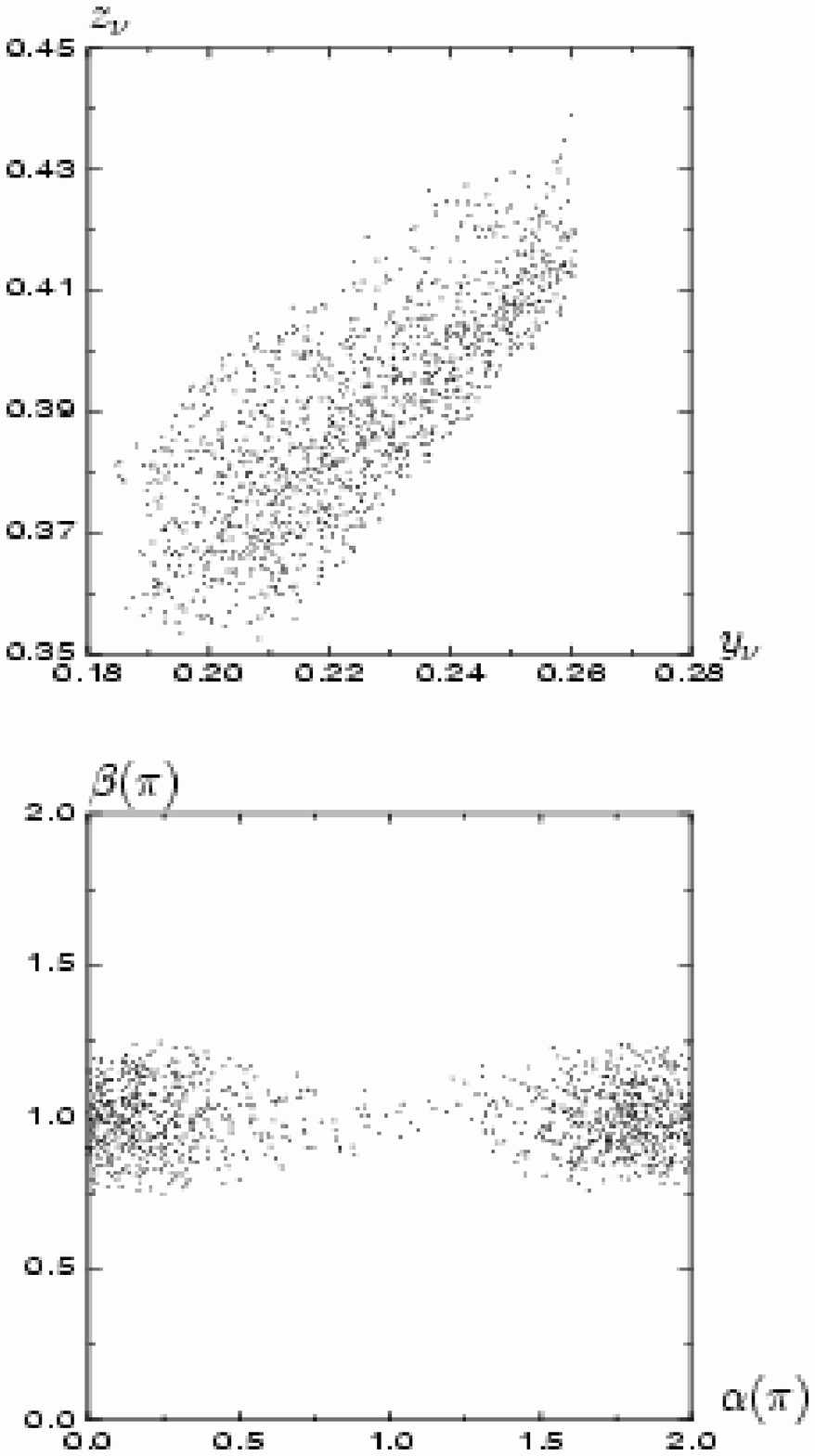,bbllx=-5.5cm,bblly=8cm,bburx=8cm,bbury=22cm,%
width=10cm,height=10cm,angle=0,clip=0} \vspace{8cm}
\caption{\underline{Non-parallel patterns of $M_l$ and $M_\nu$ in
Eq. (28a):} the parameter space of $(y^{~}_\nu, z_\nu)$ and
$(\alpha, \beta)$ at the $3\sigma$ level.}
\end{figure}
%%%%%%%%%%%%%%%%%%%%%%%%%%%%%%%%%%%%%%%%%%%

%%%%%%%%%%%%%%%%%%%% Fig. 8 %%%%%%%%%%%%%%%%
\begin{figure}[t]
\vspace{0cm}
\epsfig{file=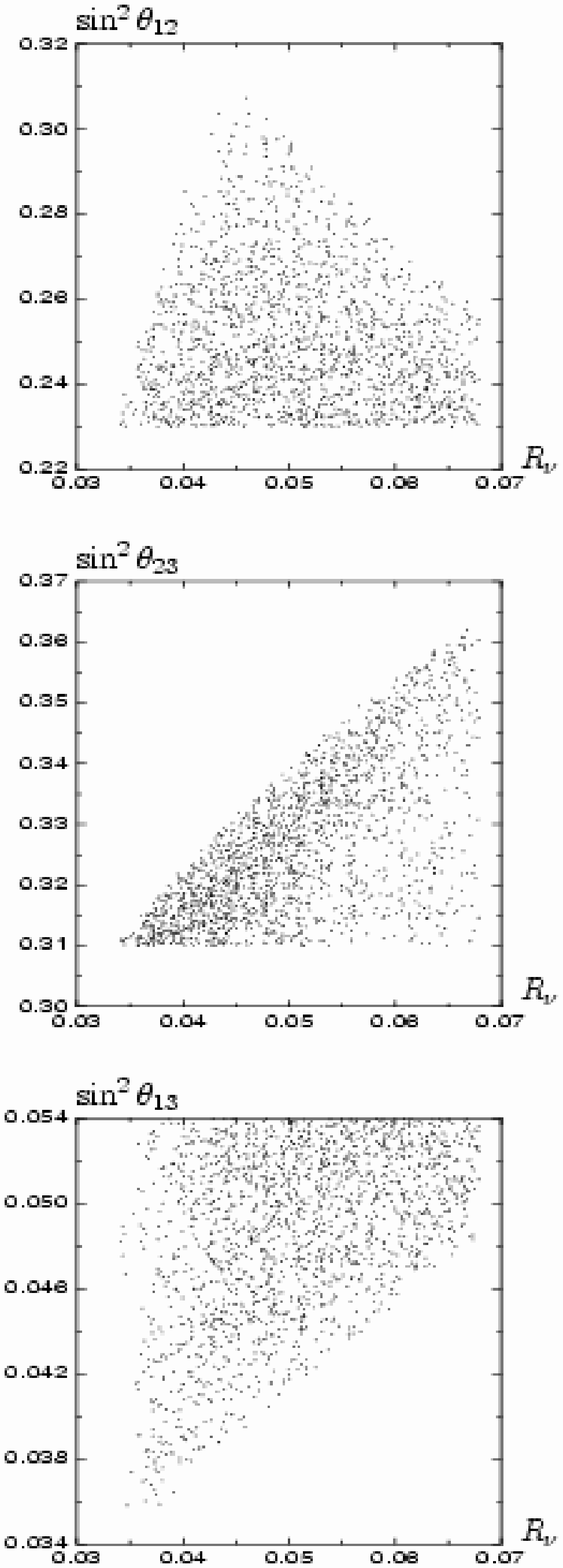,bbllx=-5cm,bblly=13cm,bburx=7cm,bbury=27cm,%
width=9cm,height=10cm,angle=0,clip=0} \vspace{11cm}
\caption{\underline{Non-parallel patterns of $M_l$ and $M_\nu$ in
Eq. (28a):} the outputs of $\sin^2 \theta_{12}$, $\sin^2
\theta_{23}$ and $\sin^2 \theta_{13}$ versus $R_\nu$ at the
$3\sigma$ level.}
\end{figure}
%%%%%%%%%%%%%%%%%%%%%%%%%%%%%%%%%%%%%%%%%%%

%%%%%%%%%%%%%%%%%%%% Fig. 9 %%%%%%%%%%%%%%%%
\begin{figure}[t]
\vspace{0cm}
\epsfig{file=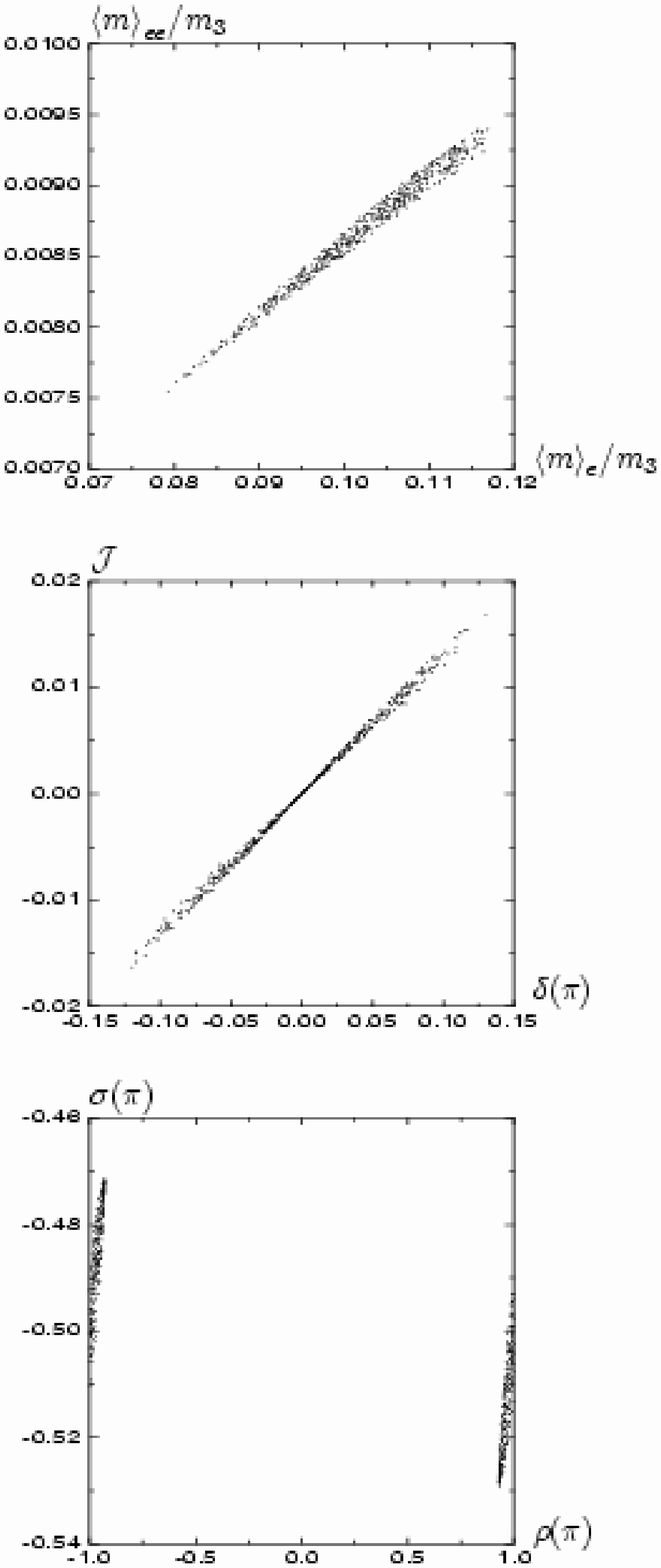,bbllx=-5cm,bblly=13cm,bburx=7cm,bbury=27cm,%
width=9cm,height=10cm,angle=0,clip=0} \vspace{11cm}
\caption{\underline{Non-parallel patterns of $M_l$ and $M_\nu$ in
Eq. (28a):} the outputs of $(\langle m\rangle_e, \langle
m\rangle_{ee})$, $(\delta, {\cal J})$ and $(\rho, \sigma)$ at the
$3\sigma$ level.}
\end{figure}
%%%%%%%%%%%%%%%%%%%%%%%%%%%%%%%%%%%%%%%%%%%

%%%%%%%%%%%%%%%%%%%% Fig. 10 %%%%%%%%%%%%%%%%
\begin{figure}[t]
\vspace{-2cm}
\epsfig{file=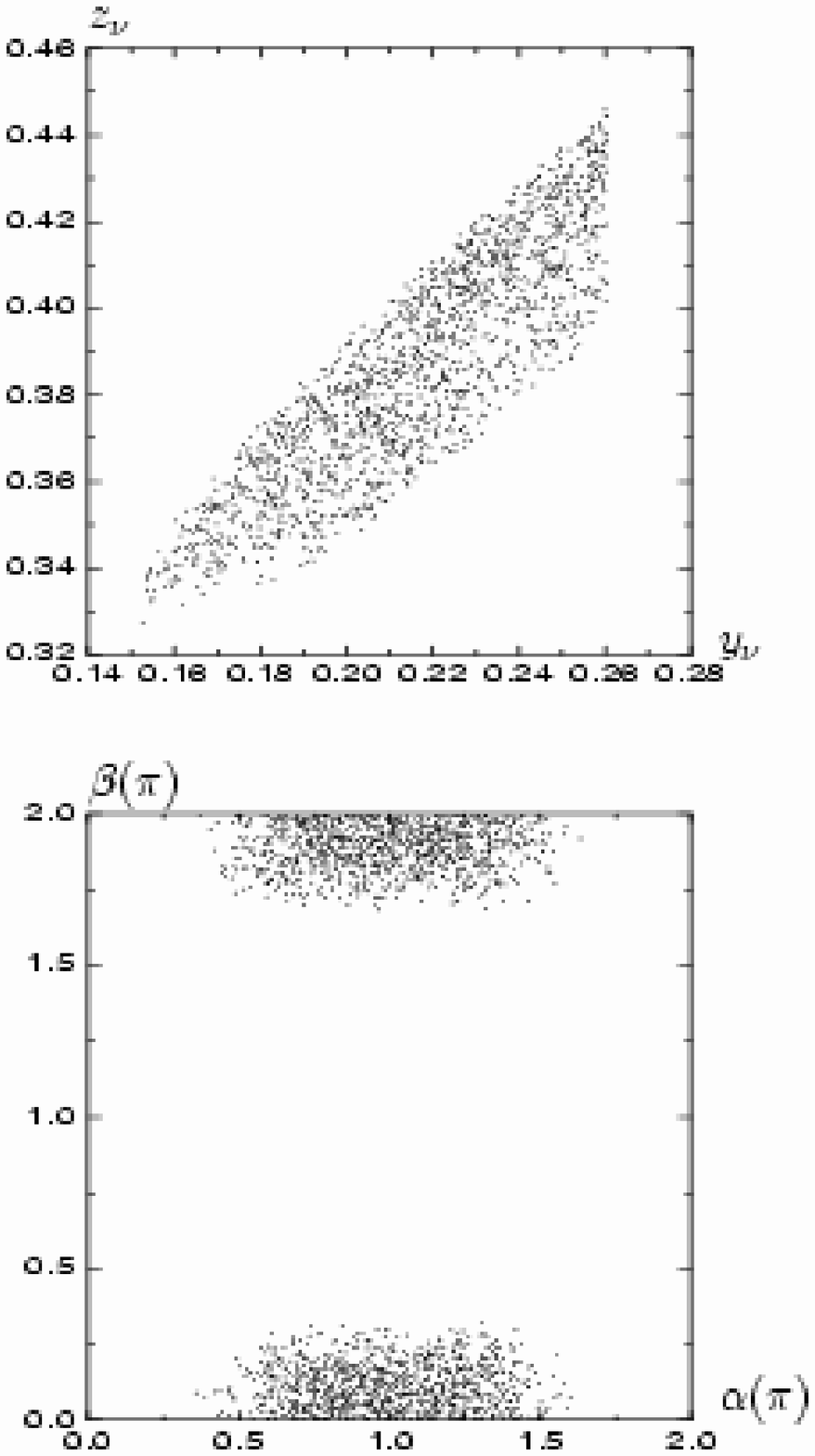,bbllx=-5.5cm,bblly=8cm,bburx=8cm,bbury=22cm,%
width=10cm,height=10cm,angle=0,clip=0} \vspace{8cm}
\caption{\underline{Non-parallel patterns of $M_l$ and $M_\nu$ in
Eq. (28c):} the parameter space of $(y^{~}_\nu, z_\nu)$ and
$(\alpha, \beta)$ at the $3\sigma$ level.}
\end{figure}
%%%%%%%%%%%%%%%%%%%%%%%%%%%%%%%%%%%%%%%%%%%

%%%%%%%%%%%%%%%%%%%% Fig. 11 %%%%%%%%%%%%%%%%
\begin{figure}[t]
\vspace{0cm}
\epsfig{file=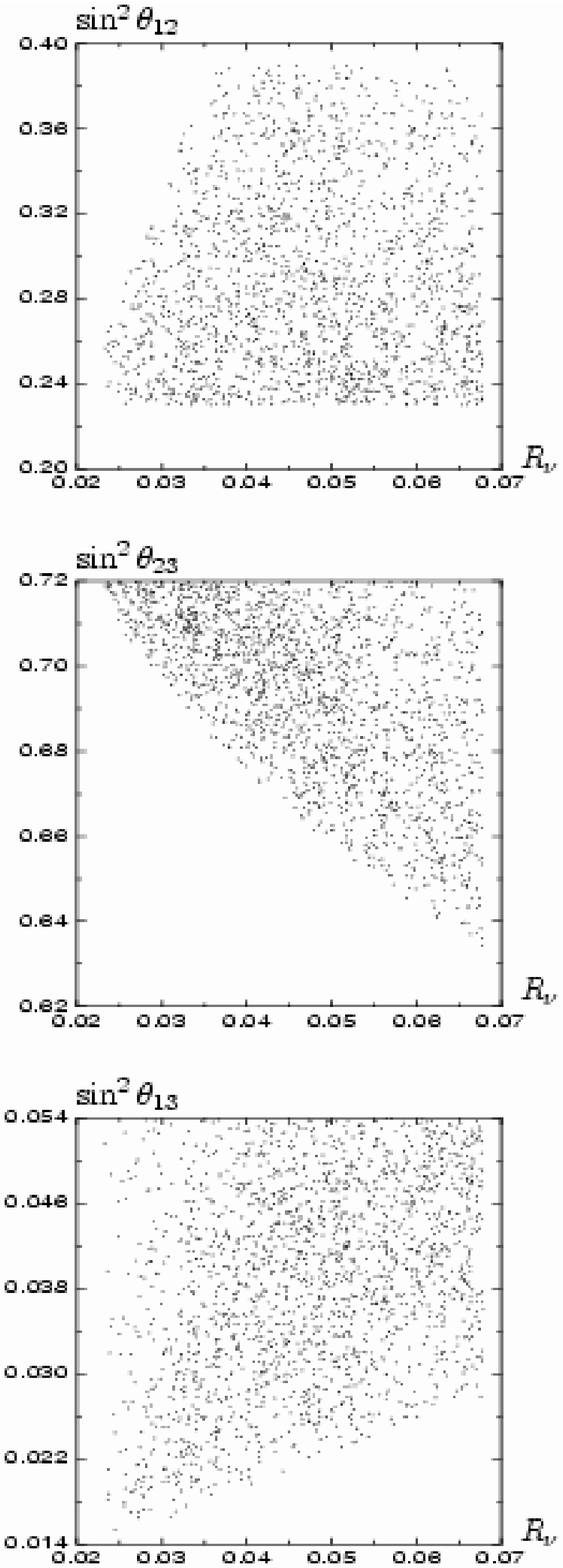,bbllx=-5cm,bblly=13cm,bburx=7cm,bbury=27cm,%
width=9cm,height=10cm,angle=0,clip=0} \vspace{11cm}
\caption{\underline{Non-parallel patterns of $M_l$ and $M_\nu$ in
Eq. (28c):} the outputs of $\sin^2 \theta_{12}$, $\sin^2
\theta_{23}$ and $\sin^2 \theta_{13}$ versus $R_\nu$ at the
$3\sigma$ level.}
\end{figure}
%%%%%%%%%%%%%%%%%%%%%%%%%%%%%%%%%%%%%%%%%%%

%%%%%%%%%%%%%%%%%%%% Fig. 12 %%%%%%%%%%%%%%%%
\begin{figure}[t]
\vspace{0cm}
\epsfig{file=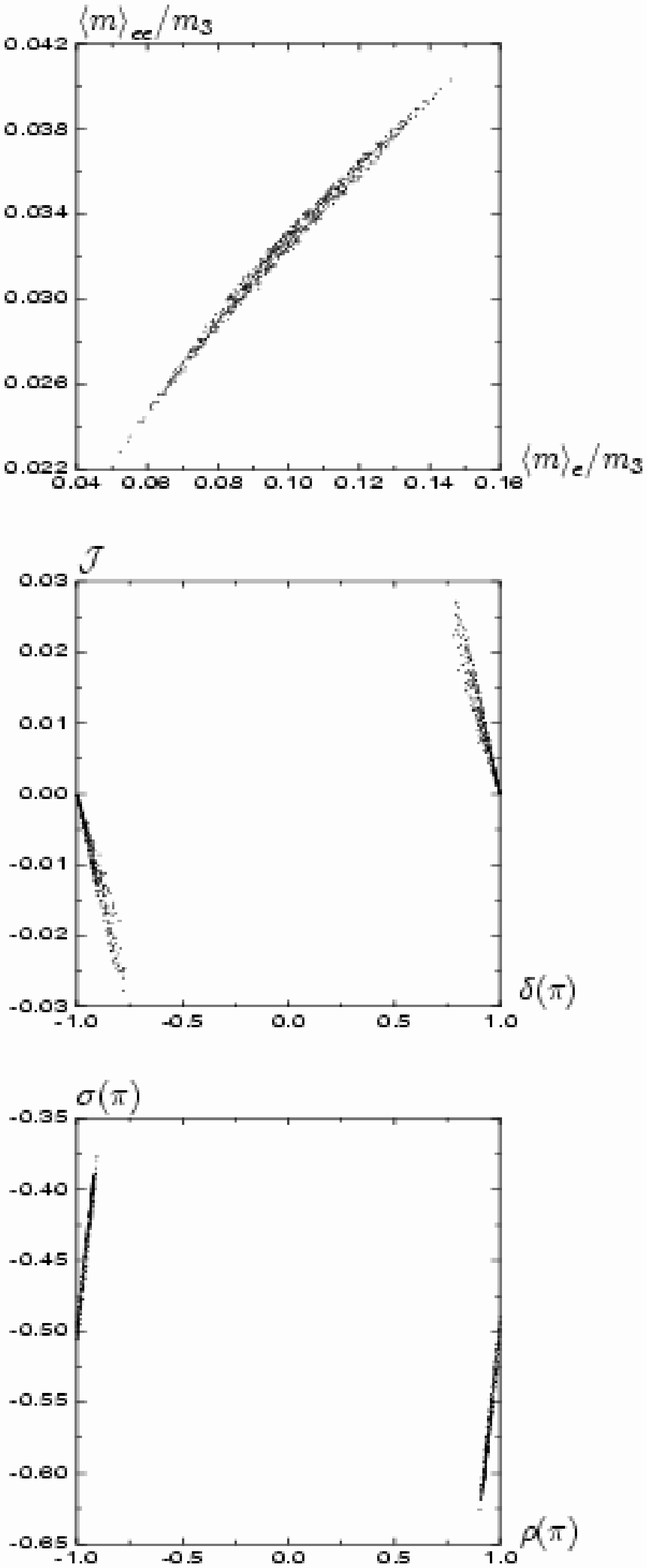,bbllx=-5cm,bblly=13cm,bburx=7cm,bbury=27cm,%
width=9cm,height=10cm,angle=0,clip=0} \vspace{11cm}
\caption{\underline{Non-parallel patterns of $M_l$ and $M_\nu$ in
Eq. (28c):} the outputs of $(\langle m\rangle_e, \langle
m\rangle_{ee})$, $(\delta, {\cal J})$ and $(\rho, \sigma)$ at the
$3\sigma$ level.}
\end{figure}
%%%%%%%%%%%%%%%%%%%%%%%%%%%%%%%%%%%%%%%%%%%

%%%%%%%%%%%%%%%%%%%% Fig. 13 %%%%%%%%%%%%%%%%
\begin{figure}[t]
\vspace{-2cm}
\epsfig{file=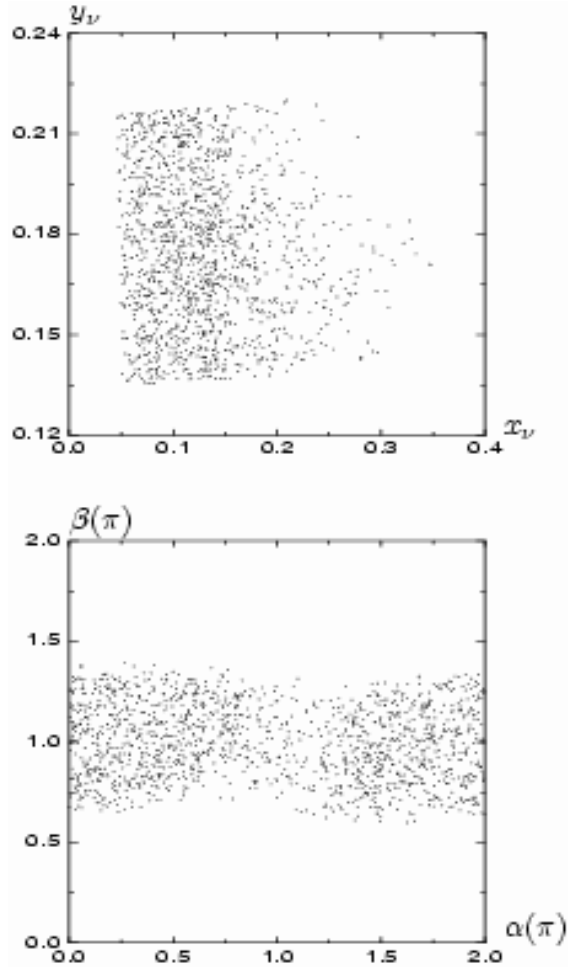,bbllx=-5.5cm,bblly=8cm,bburx=8cm,bbury=22cm,%
width=10cm,height=10cm,angle=0,clip=0} \vspace{8cm}
\caption{\underline{A simple seesaw example:} the parameter space
of $(x_\nu, y^{~}_\nu)$ and $(\alpha, \beta)$ at the $2\sigma$
level.}
\end{figure}
%%%%%%%%%%%%%%%%%%%%%%%%%%%%%%%%%%%%%%%%%%%

%%%%%%%%%%%%%%%%%%%% Fig. 14 %%%%%%%%%%%%%%%%
\begin{figure}[t]
\vspace{0cm}
\epsfig{file=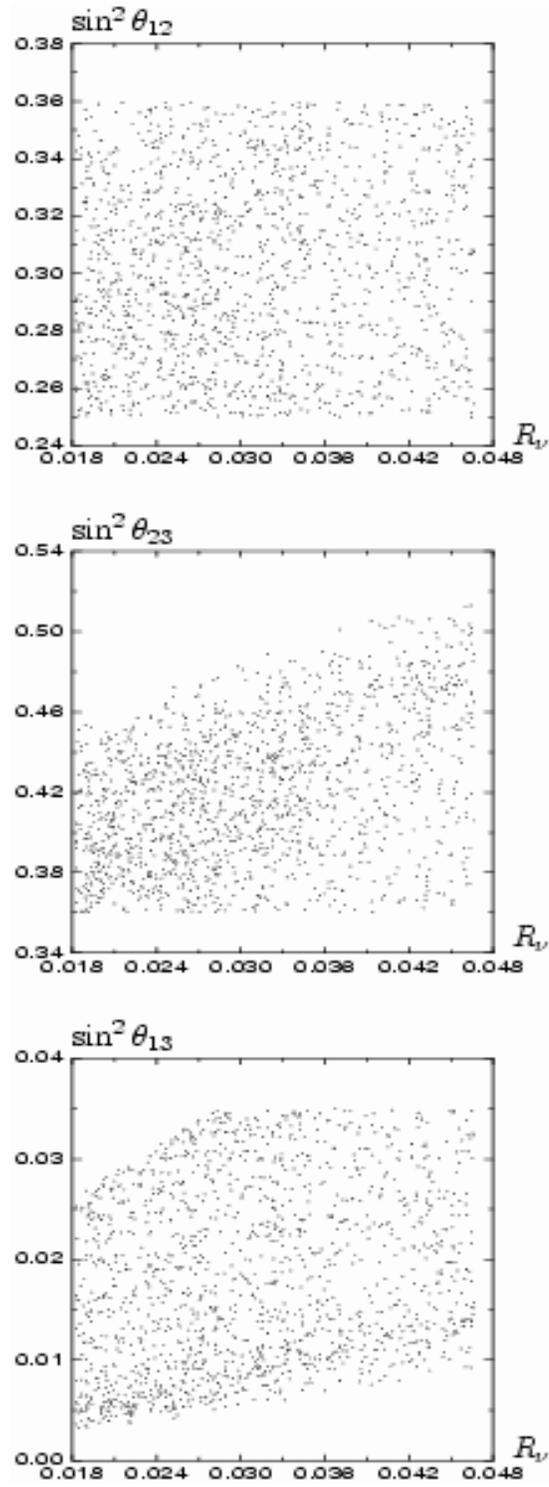,bbllx=-5cm,bblly=13cm,bburx=7cm,bbury=27cm,%
width=9cm,height=10cm,angle=0,clip=0} \vspace{11cm}
\caption{\underline{A simple seesaw example:} the outputs of
$\sin^2 \theta_{12}$, $\sin^2 \theta_{23}$ and $\sin^2
\theta_{13}$ versus $R_\nu$ at the $2\sigma$ level.}
\end{figure}
%%%%%%%%%%%%%%%%%%%%%%%%%%%%%%%%%%%%%%%%%%%

%%%%%%%%%%%%%%%%%%%% Fig. 15 %%%%%%%%%%%%%%%%
\begin{figure}[t]
\vspace{0cm}
\epsfig{file=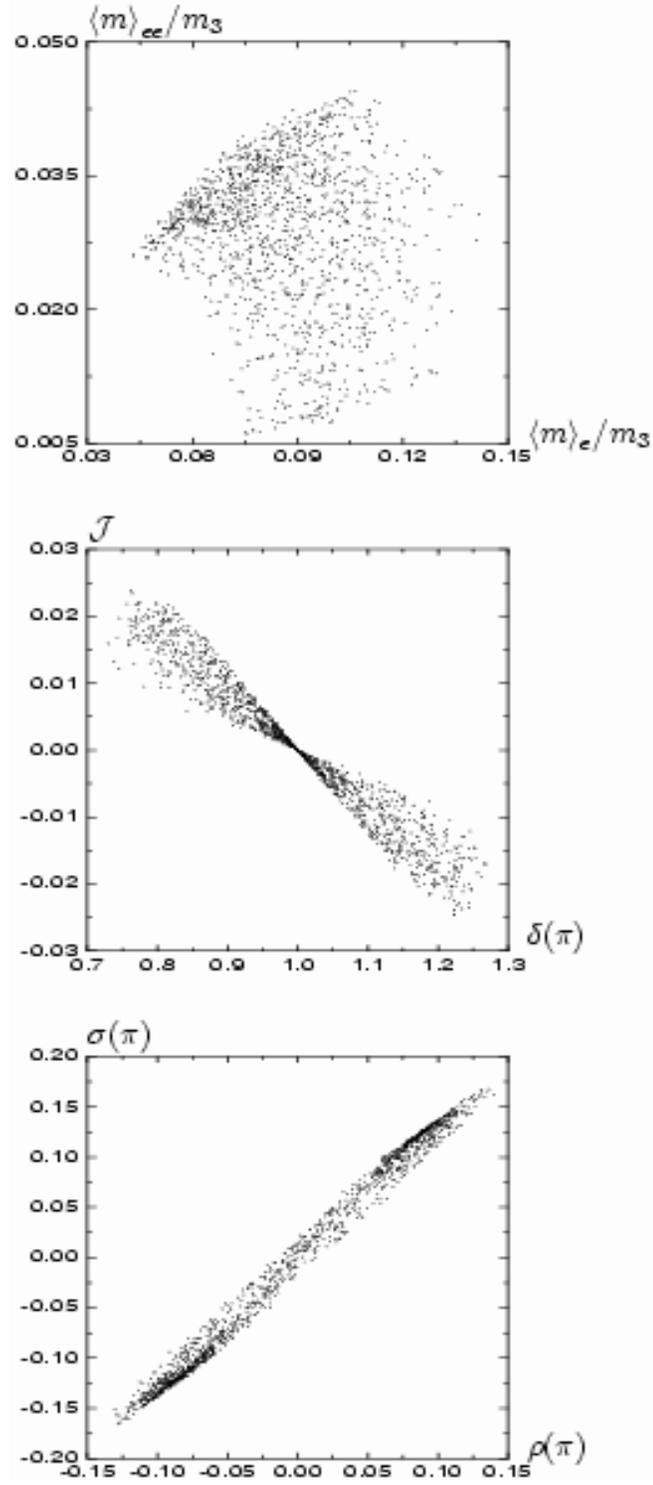,bbllx=-5cm,bblly=13cm,bburx=7cm,bbury=27cm,%
width=9cm,height=10cm,angle=0,clip=0} \vspace{11cm}
\caption{\underline{A simple seesaw example:} the outputs of
$(\langle m\rangle_e, \langle m\rangle_{ee})$, $(\delta, {\cal
J})$ and $(\rho, \sigma)$ at the $2\sigma$ level.}
\end{figure}
%%%%%%%%%%%%%%%%%%%%%%%%%%%%%%%%%%%%%%%%%%%

\end{document}